\documentclass[onecolumn,aps,prd,showpacs,eqsecnum]{revtex4}
\usepackage{amsmath,amsfonts,latexsym,epsfig}
\begin{document}
\title[Nonlinear Interactions of Gravitational Wave with Matter]{Nonlinear Interactions of Gravitational Wave with Matter in Magnetic-type
Maxwell-Vlasov Description}
\author{X.Q.Li}
\affiliation{Department of Physics, Nanchang University, Nanchang 330047, China}
\author{S.Q.Liu*}
\affiliation{Department of Physics, Nanchang University, Nanchang 330047, China}
\author{X.Y.Tao}
\affiliation{Department of Physics, Jiangxi Normal University, Nanchang, China}
\begin{abstract}
The interactions of gravitational waves with interstellar matter, dealing with
resonant wave-particle and wave-wave interactions, are considered on the basis
of magnetic-type Maxwell-Vlasov equations. It is found that the behavior of
the fields, involving the \textquotedblleft
gravitoelectromagnetic\textquotedblright\ or \textquotedblleft\ GEM
\textquotedblright\ fields, the perturbed density field and self-generated
gravitomagnetic field with low frequency, can be described by the nonlinear
coupling equations Eqs. (\ref{eq6.9})-(\ref{eq6.11}). Numerical results show
that they may collapse. In other words, due to self-condensing, a stronger GME
fields could be produced; and they could appear as the gravitational waves
with high energy reaching on Earth. In this case, Weber results, perhaps, are acceptable.
\end{abstract}
\maketitle

\section{INTRODUCTION}

It has often been noted that magnetism can be understood as the consequence of
electrostatics plus Lorentz invariance. Similarly, Newtonian gravity together
with Lorentz invariance in a consistent way must include a gravitomagnetic
field. \emph{This is the case} of gravitoelectromagnetic form of the Einstein
equations in a medium\cite{Shapiro,Bonilla,Sennovilla,Ruggiero,Iorio}. It is
well known that in the slow-motion limit of general 0relativity, accurate to
post-Newtonian (PN) order, where the field equations can be reduced as
\cite{Weinberg}:%

\begin{equation}
\nabla^{2}\phi=4\pi G\rho, \label{eq1.1}%
\end{equation}

\begin{equation}
\nabla^{2}\mathrm{\mathbf{A}}=16\pi G\rho\mathrm{\mathbf{v}}/c; \label{eq1.2}%
\end{equation}
moreover the harmonic gauge condition and the force on a unit mass reduced to%

\[
\nabla\cdot\mathrm{\mathbf{A}} + \frac{4}{c}\frac{\partial}{\partial t}\phi=
0,
\]

\[
\frac{d\mathrm{\mathbf{v}}}{dt}\approx-\nabla\phi-\frac{1}{c}\frac
{\partial\mathrm{\mathbf{A}}}{\partial t}+\frac{\mathrm{\mathbf{v}}}{c}%
\times\nabla\times\mathrm{\mathbf{A}},
\]
where $\phi$ is the Newtonian gravitational potential, $A_{i}/c=g_{i0}$ is the
mixed metric and $G$ is in the Newton's constant.

One defines the GEM fields via%

\begin{equation}
\mathrm{\mathbf{E}}_{g}=-\nabla\phi-\frac{1}{c}\frac{\partial A}{\partial
t},\quad\nabla\times\mathrm{\mathbf{A}}=\mathrm{\mathbf{B}}_{g} \label{eq1.3}%
\end{equation}
in direct analogy with electromagnetism; it follows from these definitions the
Maxwell-type field equations in the continuous
medium\cite{Braginsky,Braginsky2}%

\begin{equation}
\nabla\times\mathrm{\mathbf{B}}_{g}=\frac{16\pi}{c}\mathrm{\mathbf{j}}%
_{g}+\frac{4}{c}\frac{\partial\mathrm{\mathbf{E}}_{g}}{\partial t}%
,\ \quad{\nabla\cdot}\mathbf{D}_{g}\mathrm{\mathbf{=4\pi}}\rho_{g},
\label{eq1.4}%
\end{equation}

\begin{equation}
\nabla\times\mathrm{\mathbf{E}}_{g}=-\frac{1}{c}\frac{\partial
\mathrm{\mathbf{B}}_{g}}{\partial t},\quad{\nabla\cdot}\mathbf{B}%
_{g}\mathrm{\mathbf{=0}}\qquad\label{eq1.5}%
\end{equation}
where $\rho_{g}=-G\rho$ is \textquotedblleft matter density\textquotedblright%
\ and $\mathrm{\mathbf{j}}_{g}=-G\rho\mathrm{\mathbf{v}}=\rho_{g}%
\mathrm{\mathbf{v}}$ is \textquotedblleft matter current\textquotedblright.
Then force on a unit mass has Lorentz-type form%

\begin{equation}
\frac{d\mathrm{\mathbf{v}}}{dt}\approx\mathrm{\mathbf{E}}_{g}+\frac
{\mathrm{\mathbf{v}}}{c}\times\mathrm{\mathbf{B}}_{g}. \label{eq1.6}%
\end{equation}
And all terms of $O({v^{4}/c}^{4})$ are neglected in the above analysis. Now
one decompose the total matter density and current in Eq. (\ref{eq1.4}):
$\rho_{g}=\tilde{\rho}_{g}+\rho_{g0}$ (or $\rho=\tilde{\rho}+\rho_{0})$,
$\mathrm{\mathbf{j}}_{g}=\mathrm{\mathbf{\tilde{j}}}_{g}+\mathrm{\mathbf{j}%
}_{g0}$, in which the $\rho_{g0}$(or $\rho_{0})$ and $\mathrm{\mathbf{j}}%
_{g0}$ are the parts of a external field source in the medium, correspond to a
local additional mass disturbance, for example, by nonlinear interactions
between the fields and the medium, or some test particles. In this case it is
convenient to definite the vector quantity $\mathrm{\mathbf{D}}_{g}%
(t,\mathrm{\mathbf{r}})$ through ( in what follows one neglects the marks
\textquotedblleft\ $\sim$\textquotedblright\ in $\tilde{\rho}_{g}$and
$\mathrm{\mathbf{\tilde{j}}}_{g}$ for simplicity )%

\begin{equation}
\mathrm{\mathbf{D}}_{g}=\mathrm{\mathbf{E}}_{g}+4\pi\int\limits_{-\infty}%
^{t}{d{t}^{\prime}}\mathrm{\mathbf{j}}_{g}({t}^{\prime},\mathrm{\mathbf{r}});
\label{eq1.7}%
\end{equation}
by use of Eq.(\ref{eq1.7}) and the mass continuity equation,%

\[
\frac{\partial\rho_{\alpha}}{\partial t}+\nabla\cdot\rho_{\alpha
}\mathrm{\mathbf{v}}=0,\qquad(\alpha=g,g_{0})
\]
Eqs.(\ref{eq1.4}) and (\ref{eq1.5}) are deduced to%

\begin{equation}
\nabla\times\mathrm{\mathbf{B}}_{g}=\frac{16\pi}c\mathrm{\mathbf{j}}%
_{g0}+\frac4c\frac{\partial\mathrm{\mathbf{D}}_{g}}{\partial t},\ \quad
{\nabla\cdot}\mathrm{\mathbf{D}}_{g}\mathrm{\mathbf{=4\pi}}\rho_{g0}\ ,
\label{eq1.8}%
\end{equation}

\begin{equation}
\nabla\times\mathbf{E}_{g}=-\frac{1}{c}\frac{\partial\mathrm{\mathbf{B}}_{g}%
}{\partial t},\quad{\nabla\cdot}\mathrm{\mathbf{B}}_{g}\mathrm{\mathbf{=0\ .}}
\label{eq1.9}%
\end{equation}

A dimensional analysis of equations (\ref{eq1.1}), (\ref{eq1.3}),
(\ref{eq1.7}) and (\ref{eq1.9}) suggests that it is insightful to use the CGS
units, since these field variables have the following dimensions:%

\[
\left[  {D_{g}}\right]  \sim\left[  {E_{g}}\right]  \sim\left[  {B_{g}%
}\right]  \sim\left[  {\nabla\phi}\right]  \sim{v}/s,
\]
thus, one has%

\[
\left[  {E_{g}^{2}/G}\right]  \sim\left[  {B_{g}^{2}/G}\right]  \sim\left[
{\rho E_{g}^{2}/\rho G}\right]  \sim\left[  {\rho v^{2}}\right]  ;
\]
by substituting%

\begin{align}
\mathrm{\mathbf{E}}  &  =-\mathrm{\mathbf{E}}_{g}/4\sqrt{G},\quad
\mathrm{\mathbf{B}}=-\mathrm{\mathbf{B}}_{g}/4\sqrt{G},\quad\mathrm{\mathbf{D}%
}=-\mathrm{\mathbf{D}}_{g}/4\sqrt{G},\label{eq1.10}\\
\mathrm{\mathbf{j}}  &  =-\mathrm{\mathbf{j}}_{g}/\sqrt{G}=\sqrt{G}%
\rho\mathrm{\mathbf{v}},\quad\widehat{\rho}=-\rho_{g}/\sqrt{G}=\sqrt{G}%
\rho,\nonumber
\end{align}
the GEM field equations can be expressed in the standard form,%

\begin{equation}
\nabla\times\mathrm{\mathbf{B}}=\frac{4\pi}{c}\mathrm{\mathbf{j}}_{0}+\frac
{1}{c}\frac{\partial\mathrm{\mathbf{D}}}{\partial t},\quad\nabla
\times\mathbf{E}=-\frac{1}{c}\frac{\partial\mathrm{\mathbf{B}}}{\partial
t},\ \label{eq1.11}%
\end{equation}

\begin{equation}
\nabla\cdot\mathrm{\mathbf{D}}=\pi\hat{\rho}_{0},\quad\nabla\cdot
\mathrm{\mathbf{B}}=0; \label{eq1.12}%
\end{equation}
then Eqs.(\ref{eq1.7}) and (\ref{eq1.6}) become%

\begin{equation}
\mathrm{\mathbf{D}}=\mathrm{\mathbf{E}}+\pi\int\limits_{-\infty}^{t}%
{d{t}^{\prime}}\mathrm{\mathbf{j}}({t}^{\prime},\mathrm{\mathbf{r}}),
\label{eq1.13}%
\end{equation}

\begin{equation}
\quad\frac{d\mathrm{\mathbf{v}}}{dt}\approx-4\sqrt{G}\left[
{\mathrm{\mathbf{E}}+\frac{\mathrm{\mathbf{v}}}c\times\mathrm{\mathbf{B}}%
}\right]  . \label{eq1.14}%
\end{equation}

To consider the responses of the medium on the GEM fields, one must introduce
material relation, which describes the GEM properties of the medium. In view
of Eq.(\ref{eq1.13}), the states of the medium not only depend on a given
time-space point($t,\mathrm{\mathbf{r}})$, but also depend on previous times
and at any point of the medium. Hence, by general reasoning in physics (
independent of a specific model for the medium ) one can state that this is a
no-local linear relation in the limit of linear response ; whose Fourier
representation is%

\begin{equation}
j_{i}(\omega\mathrm{\mathbf{,k}})=\sigma_{ij}(\omega\mathrm{\mathbf{,k}}%
)E_{j}(\omega\mathrm{\mathbf{,k}}). \label{eq1.15}%
\end{equation}
In special, for common continuous medium, where the \textquotedblleft spatial
dispersion\textquotedblright\ (dependency on $\mathrm{\mathbf{k}}$) is not
important , the relation (\ref{eq1.15}) is deduced to ($\sigma_{ij}%
\rightarrow\sigma\delta_{ij})$%

\[
\mathrm{\mathbf{j}}(\omega)=\sigma(\omega)\mathrm{\mathbf{E}}(\omega),
\]
i.e. an \textquotedblleft Ohm's gravitational law\textquotedblright%
\ \cite{Ciubotariu} , where $\sigma$ is gravitational conductivity (see Refs
[10,11] for details) . Then by use of Eqs. (\ref{eq1.13}) and (\ref{eq1.15}),
one has%

\begin{equation}
\mathrm{\mathbf{D}}_{i}(\omega,\mathrm{\mathbf{k}})=\varepsilon_{ij}%
(\omega,\mathrm{\mathbf{k}})E_{j}(\omega,\mathrm{\mathbf{k}}), \label{eq1.16}%
\end{equation}
here $\varepsilon_{ij}$, called the dielectric tensor, is determined as%

\begin{equation}
\varepsilon_{ij}(\omega\mathrm{\mathbf{,k}})=\delta_{ij}+\frac{\pi i}{\omega
}\sigma_{ij}(\omega\mathrm{\mathbf{,k}})\quad(\omega\neq0). \label{eq1.17}%
\end{equation}
In the next section we shall show that the material relations
Eqs.(\ref{eq1.15}) and (\ref{eq1.16}) are relevant for GEM fields.

It has often been noted that the effects of GR can largely be understood and
treated easily within the GEM framework. For example, the famous
Lense-Thirring precession effect of GR is simpler and clearer to use the GEM
equations\cite{Hans}; and the gravitomagnetism is a useful insight for
understanding the Schiff effect\cite{Peng} and the \textquotedblleft
Faraday\textquotedblright\ effect\cite{Shapiro2}. In special, just from our
electromagnetic experience, we can infer that at a distance should present
gravitational waves predicted by GR. Therefore, the existence of the GEM
fields is equivalent of the existence of the gravitational waves predicted by
GR\cite{Peng}.

The many efforts that have been made to detect gravitational waves have so far
given no convincing evidence that they have actually been seen. In the late
1960s and early 1970s, Weber announced that he had recorded simultaneous
oscillations in detectors 1000 km apart, waves he believed originated from an
astrophysical event. But many physicists were suspicious of the results that
were several orders of magnitude higher than were theoretically predicted. In
addition, observed high energies of gravitational waves by Weber have not been
confirmed by these independent measurements. This is perhaps, due to the fact
that gravitational waves with high energies are very rarer than many
physicists had expected. On the other hand, it is highly possible that the
gravitational waves can get largely increase rate as the waves interact with
interstellar matter, so that Weber's result, perhaps, are acceptable.

In order to study the nonlinear effects on a very larger scale, where the mean
free path for collision between the particles of the matter is larger compared
with the characteristic length appearing in the problem, a kinetic treatment
is required: The physical systems should be described by the magnetic-type
Maxwell-Vlasov equations. The small amplitude and high-frequency approximation
are employed for the gravitational waves, such that perturbed techniques can
be applied. The effects deal with resonant wave-particle and wave-wave
interactions, inhomogeneties of the matter distribution and nonlinear
self-collapsing. A previous work\cite{Li} is just devoted to the study of the
nonlinear interactions in the absence of gravitomagnetic field. The present
paper considering gravitomagnetic component is a generalization of the work.
And our another work\cite{Li2} is its analog of electromagnetism.

In Section II we establish the description of collisionless kinetics and give
the linear effects. Then we present the nonlinear equations of the fields with
low frequency and high frequency, starting from GME equations in Sections III
and IV . In Section V we focus on the motion of matter disturbed by GME
fields. As a result, the nonlinear controlling equations, involving the
coupling of GEM fields with the perturbed density field and self-generated
gravitomagnetic field, are presented in Section VI . A numerical integral of
the controlling equations is given in Section VII. Finally we sketch the
conclusions, stressing the possibility of detecting gravitational waves.

\section{KINETIC DESCRIPTION AND LINEAR EFFECTS}

Taking into account the presence of massive dark matter in the universe, one
should treat a two-component self-gravitating system . The collisionless
Boltzman equations for the distribution function $f_{\alpha}$are\cite{Li}%

\begin{equation}
\frac{\partial f_{\alpha}}{\partial t}+{\mathrm{\mathbf{v}}}\frac{\partial
f_{\alpha}}{\partial{\mathrm{\mathbf{r}}}}+({\mathrm{\mathbf{a+F}}})\cdot
\frac{\partial f_{\alpha}}{\partial{\mathrm{\mathbf{p}}}}=0\quad(\alpha=1,2),
\label{eq2.1}%
\end{equation}
where $\mathrm{\mathbf{a}}$ is the non-gravitational term, $\mathrm{\mathbf{F}%
}$ is Lorentz-type force:%

\begin{equation}
\mathrm{\mathbf{F}}=\frac{d\mathrm{\mathbf{p}}}{dt}\approx q_{\alpha}\left[
{\mathrm{\mathbf{E}}+\frac{\mathrm{\mathbf{v}}}{c}\times\mathrm{\mathbf{B}}%
}\right]  \quad, \label{eq2.2}%
\end{equation}
In which gravitoelectric field $\mathrm{\mathbf{E}}$ and gravitomagnetic field
$\mathrm{\mathbf{B}}$ satisfy Eqs.(\ref{eq1.11}) and(\ref{eq1.12}), and the
\textquotedblleft charge\textquotedblright\ is%

\[
q_{\alpha}=-4\sqrt{G}m_{\alpha}.
\]
The density $n_{\alpha\text{ }}$and the current density $\mathrm{\mathbf{j}}$
are connected with the particle distribution through%

\begin{equation}
n_{a}(\mathrm{\mathbf{r}},t)=\int{f_{\alpha}}(\mathrm{\mathbf{r}%
},\mathrm{\mathbf{v}},t)\frac{d\mathrm{\mathbf{p}}}{(2\pi)^{3}}, \label{eq2.3}%
\end{equation}
and%

\begin{equation}
\mathrm{\mathbf{j}}(\mathrm{\mathbf{r}},t)=-\frac{1}{4}\sum\limits_{\alpha
}{\int{q_{\alpha}\mathrm{\mathbf{v}}}}f_{\alpha}(\mathrm{\mathbf{r}%
},\mathrm{\mathbf{v}},t)\frac{d\mathrm{\mathbf{p}}}{(2\pi)^{3}}. \label{eq2.4}%
\end{equation}
We can assume to the gravitation effects%

\begin{equation}
\rho_{2}=n_{0}m_{2}\ll\rho_{2}=n_{0}m_{2}, \label{eq2.5}%
\end{equation}
where $\rho_{1}$ denotes the density of bright matter, and $\rho_{2}$ the
density of dark matter. We divide $f_{\alpha}$, $\mathrm{\mathbf{E}}$ and
$\mathrm{\mathbf{B}}$ into two parts: unperturbed and perturbed parts,%

\begin{equation}
f_{\alpha}=f_{\alpha}^{R}+f_{\alpha}^{T},\quad\mathrm{\mathbf{E}%
}=\mathrm{\mathbf{E}}^{R}+\mathrm{\mathbf{E}}{}^{T},\quad\mathrm{\mathbf{B}%
}=\mathrm{\mathbf{B}}^{R}+\mathrm{\mathbf{B}}{}^{T}\ . \label{eq2.6}%
\end{equation}
As the $f_{\alpha}$ is closely coupling with GEM, the perturbed distribution
can be expanded in power%

\begin{equation}
f_{\alpha}^{T}=\sum\limits_{i}f^{T(i)} \label{eq2.7a}%
\end{equation}
provided that the perturbed field $E^{T}$ is weak, i.e.%

\begin{equation}
\bar{W}=\frac{|\mathrm{\mathbf{E}}^{T}|^{2}}{8\pi n_{0}T_{0}}<<1,
\label{eq2.7b}%
\end{equation}
where the index $i$ indicates the $i$-th power of $E^{T}$. In this case that
equation for the unperturbed state%

\begin{equation}
\frac{\partial f_{\alpha}^{R}}{\partial t}+\mathrm{\mathbf{v}}\cdot
\frac{\partial f_{\alpha}^{R}}{\partial\mathrm{\mathbf{r}}}%
+(\mathrm{\mathbf{a}}+\mathrm{\mathbf{F}}^{R})\cdot\frac{\partial f_{\alpha
}^{R}}{\partial\mathrm{\mathbf{p}}}=0 \label{eq2.8}%
\end{equation}
reduced to%

\[
\frac{\partial f_{\alpha}^{R}}{\partial t}=0,
\]
with a relevant solution%

\begin{equation}
f_{\alpha,p_{x}}^{R}{}\equiv\int{f_{\alpha}^{R}\frac{dp_{y}dp_{z}}{(2\pi)^{2}%
}}=\frac{(2\pi)^{1/2}}{(m_{\alpha}v_{T\alpha})}n_{0}e^{-\frac{p_{x}^{2}%
}{2m_{\alpha}^{2}v_{T\alpha}^{2}}}. \label{eq2.9}%
\end{equation}
Substituting Eq.(\ref{eq2.6}) into Eq.(\ref{eq2.1}) and subtracting
Eq.(\ref{eq2.8}), putting $\mathrm{\mathbf{a}}+\mathrm{\mathbf{g}}^{R}=0,$ yield%

\begin{equation}
\frac{\partial f_{\alpha}^{T}}{\partial t}+\mathrm{\mathbf{v}}\cdot
\frac{\partial f_{\alpha}^{T}}{\partial\mathrm{\mathbf{r}}}+\mathrm{\mathbf{F}%
}^{T}\cdot\frac{\partial f_{\alpha}^{R}}{\partial\mathrm{\mathbf{p}}%
}+\mathrm{\mathbf{F}}^{T}\cdot\frac{\partial f_{\alpha}^{T}}{\partial
\mathrm{\mathbf{p}}}=0. \label{eq2.10}%
\end{equation}
Substituting Eq.(\ref{eq2.7a}) into Eq.(\ref{eq2.10}) and expanding
$A=(\mathrm{\mathbf{F}}^{T},f_{\alpha})$in a Fourier series%

\begin{equation}
A(\mathrm{\mathbf{r}},\mathrm{\mathbf{v}},t)=\int{A_{k}}e^{-i\omega
t+i\mathrm{\mathbf{k}}\cdot\mathrm{\mathbf{r}}}dk, \label{eq2.11}%
\end{equation}

\[
A_{k}\equiv A_{\mathrm{\mathbf{k}},\omega},dk=d\mathrm{\mathbf{k}}d\omega,
\]
we get from Eq.(\ref{eq2.10}):%

\begin{equation}
i(\omega-\mathrm{\mathbf{k}}\cdot\mathrm{\mathbf{v}})f_{\alpha,k}^{T(1)}%
=\int{\mathrm{\mathbf{F}}_{k}^{T}}\cdot\frac{\partial f_{\alpha,k_{2}}^{R}%
}{\partial\mathrm{\mathbf{p}}}\delta(k-k_{1}-k_{2})dk_{1}dk_{2},
\label{eq2.12}%
\end{equation}

\begin{equation}
i(\omega-\mathrm{\mathbf{k}}\cdot\mathrm{\mathbf{v}})f_{\alpha,k}^{T(2)}%
=\int{\mathrm{\mathbf{F}}_{k_{1}}^{T}}\cdot\frac{\partial f_{\alpha,k_{2}%
}^{T(1)}}{\partial\mathrm{\mathbf{p}}}\delta(k-k_{1}-k_{2})dk_{1}dk_{2},
\label{eq2.13}%
\end{equation}

\begin{equation}
i(\omega-\mathrm{\mathbf{k}}\cdot\mathrm{\mathbf{v}})f_{\alpha,k}^{T(3)}%
=\int{\mathrm{\mathbf{F}}_{k_{1}}^{T}}\cdot\frac{\partial f_{\alpha,k_{2}%
}^{T(2)}}{\partial\mathrm{\mathbf{p}}}\delta(k-k_{1}-k_{2})dk_{1}dk_{2}.
\label{eq2.14}%
\end{equation}
Dividing $\mathrm{\mathbf{j}}$ into to two parts,%

\begin{equation}
\mathrm{\mathbf{j}}=\mathrm{\mathbf{j}}^{R}+\mathrm{\mathbf{j}}^{T},
\label{eq2.15}%
\end{equation}
and expanding $\mathrm{\mathbf{j}}^{T}$ in powers of $E^{T}$%

\begin{equation}
\mathrm{\mathbf{j}}^{T(i)}=-\frac14\sum\limits_{\alpha}{\int{q_{\alpha
}\mathrm{\mathbf{v}}}}f_{\alpha}^{T(i)}\frac{d\mathrm{\mathbf{p}}}{(2\pi)^{3}%
}, \label{eq2.16}%
\end{equation}
one can obtain the linear current from Eqs.(\ref{eq2.12}), (\ref{eq2.9}), and
(\ref{eq2.16})%

\begin{equation}
j_{k,i}^{T}=\sigma_{ij}(\omega,\mathrm{\mathbf{k}})E_{k,j}^{T}\ ,
\label{eq2.17}%
\end{equation}
This is the relation Eq.(\ref{eq1.15}); where%

\begin{equation}
\sigma_{ij}(\omega,\mathrm{\mathbf{k}})=-\frac{1}{4}\sum\limits_{\alpha}%
{\int{\frac{v_{i}q_{\alpha}^{2}\left[  {\delta_{js}(1-\frac{\mathrm{\mathbf{k}%
}\cdot\mathrm{\mathbf{v}}}{\omega})+\frac{k_{s}v_{j}}{\omega}}\right]
}{i(\omega-\mathrm{\mathbf{k}}\cdot\mathrm{\mathbf{v}}+i\varepsilon)}}}%
\frac{\partial f_{\alpha}^{R}}{\partial p_{s}}\frac{d\mathrm{\mathbf{p}}%
}{(2\pi)^{3}}\quad; \label{eq2.18}%
\end{equation}
and here we have into account the slow change in $f_{\alpha}^{R}$, i.e.
$f_{\alpha,k_{2}}^{R}\approx f_{\alpha}^{R}\delta(k_{2})$ ; the term
$i\varepsilon$ arises from the Landau rule.

Similarly, we get the nonlinear currents (up to the third order) from
Eqs.(\ref{eq2.13}) and (\ref{eq2.14})%

\begin{equation}
\mathrm{\mathbf{j}}_{k}^{\left(  2\right)  }=\sum\limits_{\alpha}%
{\int{\mathrm{\mathbf{S}}_{k,k_{2,}k_{3}}^{\alpha}E_{k_{1}}^{T}E_{k_{2}}%
^{T}dk_{1}dk_{2}\delta\left(  {k-k_{1}-k_{2}}\right)  }}, \label{eq2.19}%
\end{equation}

\begin{equation}
\mathrm{\mathbf{j}}_{k}^{(3)}=\sum\limits_{\alpha}{\int{\mathrm{\mathbf{G}%
}_{k,k_{1},k_{2},k_{3}}^{\alpha}E_{k_{1}}^{T}E_{k_{2}}^{T}E_{k_{3}}^{T}%
\delta\left(  {k-k_{1}-k_{2}-k_{3}}\right)  dk_{1}dk_{2}dk_{3}}}
\label{eq2.20}%
\end{equation}
with%

\begin{equation}
\mathrm{\mathbf{S}}_{k,k_{1},k_{2}}^{\alpha}=\frac14q_{\alpha}^{3}\int
{\frac{\mathrm{\mathbf{v}}\left(  {\mathrm{\mathbf{\tilde{e}}}%
_{\mathrm{\mathbf{k}}_{1}}^{\sigma}\cdot\frac\partial{\partial
\mathrm{\mathbf{p}}}}\right)  }{\left(  {\omega-\mathrm{\mathbf{k}}%
\cdot\mathrm{\mathbf{v}}+i\varepsilon}\right)  }\frac{\left(
{\mathrm{\mathbf{e}}_{\mathrm{\mathbf{k}}_{2}}^{\sigma}\cdot\frac
\partial{\partial\mathrm{\mathbf{p}}}}\right)  }{\left(  {\omega
_{2}-\mathrm{\mathbf{k}}_{2}\cdot\mathrm{\mathbf{v}}+i\varepsilon}\right)
}f_{\alpha}^{R}\frac{d\mathrm{\mathbf{p}}}{\left(  {2\pi}\right)  ^{3}}},
\label{eq2.21}%
\end{equation}

\begin{align}
\mathrm{\mathbf{G}}_{k,k_{1},k_{2},k_{3}}^{\alpha}  &  =-\frac{1}{4}%
iq_{\alpha}^{4}\int{\frac{\mathrm{\mathbf{v}}\cdot d\mathrm{\mathbf{p}}}%
{(2\pi)^{3}}\frac{1}{\left(  {\omega-\mathrm{\mathbf{k}}\cdot
\mathrm{\mathbf{v}}+i\varepsilon}\right)  }\left(  {\mathrm{\mathbf{\tilde{e}%
}}_{\mathrm{\mathbf{k}}}^{\sigma}\cdot\frac{\partial}{\partial
\mathrm{\mathbf{p}}}}\right)  \frac{1}{\left[  {\left(  {\omega-\omega_{1}%
}\right)  -\left(  {\mathrm{\mathbf{k}}-\mathrm{\mathbf{k}}_{1}}\right)
\cdot\mathrm{\mathbf{v}}+i\varepsilon}\right]  }}\label{eq2.22}\\
&  \left(  {\mathrm{\mathbf{\tilde{e}}}_{\mathrm{\mathbf{k}}_{2}}^{\sigma
}\cdot\frac{\partial}{\partial\mathrm{\mathbf{p}}}}\right)  \frac{1}{\left(
{\omega_{3}-\mathrm{\mathbf{k}}_{\mathrm{\mathbf{3}}}\cdot\mathrm{\mathbf{v}%
}+i\varepsilon}\right)  }\left(  {\mathrm{\mathbf{e}}_{\mathrm{\mathbf{k}}%
_{3}}^{\sigma}\cdot\frac{\partial}{\partial\mathrm{\mathbf{p}}}}\right)
f_{\alpha}^{R}.\nonumber
\end{align}
On the other hand, we can obtain the field equation from the Maxwell--type GEM equations%
\begin{equation}
\left(
{k^{2}-\frac{\omega^{2}}{c^{2}}\varepsilon_{k}^{\sigma}}\right)
E_{k}^{T\sigma}=\frac{4\pi
i}{c^{2}}\omega(\mathrm{\mathbf{e}}_{k}^{\sigma *}\cdot
\sum\limits_{{n \ge 2}}{\mathrm{\mathbf{j}}_{k}^{T(n)}}),
\label{eq2.23}%
\end{equation}
where%

\begin{equation}
\varepsilon_{k}^{\sigma}\equiv\varepsilon_{\omega,\mathrm{\mathbf{k}}}%
^{\sigma}=\varepsilon_{ij}^{\sigma}(\omega,\mathrm{\mathbf{k}}%
)e_{\mathrm{\mathbf{k}},i}^{\sigma}e_{\mathrm{\mathbf{k}},j}^{\sigma*}%
+\frac{c^{2}}{\omega^{2}}(\mathrm{\mathbf{k}}\cdot\mathrm{\mathbf{e}%
}_{\mathrm{\mathbf{k}}}^{\sigma})(\mathrm{\mathbf{k}}\cdot\mathrm{\mathbf{e}%
}_{\mathrm{\mathbf{k}}}^{\sigma*}) \label{eq2.24}%
\end{equation}
is ``dielectric constant'' for $\sigma$ mode (longitudinal mode, $\sigma=l$
and transverse $\sigma=t)$ and $\mathrm{\mathbf{E}}_{k}^{^{T\sigma}}%
=E_{k}^{^{T\sigma}}\mathrm{\mathbf{e}}_{\mathrm{\mathbf{k}}}^{\sigma}$ with
$\mathrm{\mathbf{e}}_{\mathrm{\mathbf{k}},i}^{\sigma}\mathrm{\mathbf{e}%
}_{\mathrm{\mathbf{k}},i}^{\sigma*}=1$. Using Eqs.(\ref{eq1.17}),
(\ref{eq2.17})and (\ref{eq2.18}), one has%

\begin{equation}
\varepsilon_{k}^{l}\equiv\varepsilon_{ij}^{\sigma}(\omega,\mathrm{\mathbf{k}%
})\frac{k_{i}k_{j}}{k^{2}}=1-\sum\limits_{\alpha}{\frac1{k^{2}}\frac
{\omega_{p\alpha}^{2}}{v_{T\alpha}^{2}}}\left[  {1-Z}\left(  {\frac
\omega{\sqrt{2}kv_{T\alpha}}}\right)  \right]  , \label{eq2.25}%
\end{equation}

\begin{equation}
\varepsilon_{k}^{t}\equiv\varepsilon_{ij}^{\sigma}(\omega,\mathrm{\mathbf{k}%
})e_{\mathrm{\mathbf{k}},i}^{\sigma}e_{\mathrm{\mathbf{k}},j}^{\sigma*}%
=1+\sum\limits_{\alpha}{\frac1{k^{2}}\frac{\omega_{p\alpha}^{2}}\omega
Z}\left(  {\frac\omega{\sqrt{2}kv_{T\alpha}}}\right)  \label{eq2.26}%
\end{equation}
with%

\begin{equation}
\omega_{p\alpha}^{2}=\frac{\pi q_{\alpha}^{2}n_{0}}{4m_{\alpha}}=4\pi
G\rho_{\alpha}, \label{eq2.27}%
\end{equation}
where%

\begin{equation}
Z\left(  \frac{\omega}{\sqrt{2}kv_{T1}}\right)  \equiv\quad\int_{-\infty
}^{\infty}{\frac{x/\sqrt{\pi}}{x-\xi+i\varepsilon}e^{-\xi^{2}}d\xi}
\label{eq2.28}%
\end{equation}
is dispersion function\cite{Li2}:%

\begin{equation}
Z(x)\approx1+\frac1{2x^{2}}+\frac3{4x^{4}}-i\sqrt{\pi}xe^{-x^{2}},\quad x\gg1,
\label{eq2.29}%
\end{equation}

\begin{equation}
Z(x)\approx2x^{2}-i\sqrt{\pi}xe^{-x^{2}}\approx2x^{2}-i\sqrt{\pi}x,{\quad
x\ll1.} \label{eq2.30}%
\end{equation}
For high-frequency field%

\begin{equation}
\omega\gg kv_{T1}\gg kv_{T2}, \label{eq2.31}%
\end{equation}
i.e.$x\gg1$, therefore, neglecting the damping term, one obtain from
Eqs.(\ref{eq2.25}) and (\ref{eq2.26})%

\begin{equation}
\varepsilon_{k}^{l}=1+\frac{\omega_{p2}^{2}}{\omega^{2}}+\frac{\omega_{p1}%
^{2}}{\omega^{2}}+\frac{\omega_{p2}^{2}}{\omega^{2}}\frac{3k^{2}v_{T2}^{2}%
}{\omega^{2}}+\frac{\omega_{p1}^{2}}{\omega^{2}}\frac{3k^{2}v_{T1}^{2}}%
{\omega^{4}} \label{eq2.32}%
\end{equation}
and%

\begin{equation}
\varepsilon_{k}^{t}\approx1+\frac{\omega_{p1}^{2}}{\omega^{2}}+\frac
{\omega_{p2}^{2}}{\omega^{2}}. \label{eq2.33}%
\end{equation}
For the low-frequency fields, the following conditions are met%

\begin{equation}
v_{T2}\gg\omega^{\prime}/k^{\prime}\ll v_{T1}\quad,\quad\omega^{\prime}%
\ll\omega_{p1}, \label{eq2.34}%
\end{equation}
Then%

\begin{equation}
\varepsilon_{k^{\prime}}^{l}{}=1+(\varepsilon_{k^{\prime}}^{1(l)}%
{}-1)+(\varepsilon_{k^{\prime}}^{2(l)}{}-1)\approx-\frac{\omega_{p1}^{2}%
}{k^{\prime}{}^{2}v_{T1}^{2}}-\frac{\omega_{p2}^{2}}{k^{\prime}{}^{2}%
v_{T2}^{2}}, \label{eq2.35}%
\end{equation}

\begin{equation}
\varepsilon_{k^{\prime}}^{t}\approx1+\frac{\omega_{p1}^{2}}{k^{\prime}{}%
^{2}v_{T1}^{2}}+\frac{\omega_{p2}^{2}}{k^{\prime}{}^{2}v_{T2}^{2}}%
-i\sqrt{\frac{\pi}{2}}\frac{\omega_{p2}}{\omega^{\prime}}\frac{\omega_{p2}%
}{k^{\prime}v_{T2}}. \label{eq2.36}%
\end{equation}

Taking Eq.(\ref{eq2.33}) into consideration, one gets the linear dispersion
relationship from Eq.$(\ref{eq2.23})$ for the transverse oscillation with high-frequency:%

\begin{equation}
\omega^{2}=k^{2}c^{2}-\left(  {\omega_{p1}^{2}+\omega_{p2}^{2}}\right)  .
\label{eq2.37}%
\end{equation}
Usually, there is not steady longitudinal high-frequency mode on basis of
Eq.(\ref{eq2.32}), which consistent with the analyses for the
electro-gravitation kinetics\cite{Li}. Consider that only the oscillations
with frequency is close to the proper frequency of the medium, which is
similar to plasma case\cite{Zakharov}; this means%

\begin{equation}
\omega^{2}\simeq\omega_{pn}^{2}+\tilde{k}^{2}c^{2}\approx\omega_{p2}^{2}%
,\quad\left(  \omega_{pn}^{2}\equiv\omega_{p1}^{2}+\omega_{p2}^{2}\gg\tilde
{k}^{2}c^{2}\right)  . \label{eq2.38}%
\end{equation}
Eq.(\ref{eq2.38}) is a branch of dispersion relationship [Eq.(\ref{eq2.37})].
In what follows we omit the mark `$\sim$' in Eq.(\ref{eq2.38}) for simplicity.

\section{LOW-FREQUENCY TRANSVERSE FIELD EQUATION}

First of all, let us study the case of the low-frequency transverse field
$E_{k}^{T\sigma}=E_{k}^{Tt}\equiv E_{k}^{TS}$ in Eq.(\ref{eq2.23}%
).\emph{\ }For the second order nonlinear current, because of delta function
in Eq.(\ref{eq2.19}), it must met $\mathrm{\mathbf{k}}=\mathrm{\mathbf{k}}%
_{1}+\mathrm{\mathbf{k}}_{2},\omega=\omega_{1}+\omega_{2}$; since $\left(
{\mathrm{\mathbf{k}},\omega}\right)  $ belong to low-frequency wave, then
$\omega_{1},\omega_{2}$ must be high-frequency and are of opposite sign. In
this case, the second order term in Eq.(\ref{eq2.19}) becomes: $\left[
{E_{k_{1}}^{T\left(  +\right)  }E_{k_{2}}^{T\left(  -\right)  }+E_{k_{1}%
}^{T\left(  -\right)  }E_{k_{2}}^{T\left(  +\right)  }}\right]  $, where upper
indices \textquotedblleft+\textquotedblright\ and \textquotedblright%
-\textquotedblleft\ denote the positive and negative frequency parts of
high-frequency perturbations, respectively. Hence%

\begin{equation}
\mathrm{\mathbf{j}}_{k}^{\left(  2\right)  }=\sum\limits_{\alpha}%
{\int{\mathrm{\mathbf{S}}_{k,k_{1},k_{2}}^{\alpha(t)}\left(  {E_{k_{1}%
}^{T\left(  +\right)  }E_{k_{2}}^{T\left(  -\right)  }+E_{k_{1}}^{T\left(
-\right)  }E_{k_{2}}^{T\left(  +\right)  }}\right)  \delta\left(
{k-k_{1}-k_{2}}\right)  dk_{1}dk_{2},}} \label{eq3.1}%
\end{equation}
where%

\begin{equation}
\mathrm{\mathbf{S}}_{k,k_{1},k_{2}}^{\alpha(t)}=\frac14q_{\alpha}^{3}%
\int{\mathrm{\mathbf{v}}\frac{\mathrm{\mathbf{\tilde{e}}}_{\mathrm{\mathbf{k}%
}_{\mathrm{\mathbf{1}}}}^{t}\cdot\frac\partial{\partial\mathrm{\mathbf{p}}}%
}{\left(  {\omega-\mathrm{\mathbf{k}}\cdot\mathrm{\mathbf{v}}+i\varepsilon
}\right)  }\frac{\mathrm{\mathbf{e}}_{\mathrm{\mathbf{k}}_{2}}^{t}\cdot
\frac\partial{\partial\mathrm{\mathbf{p}}}f_{\alpha}^{R}}{\omega
_{2}-\mathrm{\mathbf{k}}_{\mathrm{\mathbf{2}}}\cdot\mathrm{\mathbf{v}%
}+i\varepsilon}}\frac{\mathrm{\mathbf{dp}}}{\left(  2{\pi}\right)  ^{3}}\quad.
\label{eq3.2}%
\end{equation}
Using the substitution $k_{1}\to k_{2},k_{2}\to k_{1}$ in the second integral
term of Eq.(\ref{eq3.1}), it yields%

\begin{equation}
\mathrm{\mathbf{j}}_{k}^{\left(  2\right)  }=\sum\limits_{\alpha}\int{\left(
{\mathrm{\mathbf{S}}_{k,k_{1},k_{2}}^{\alpha(t)}+\mathrm{\mathbf{S}}%
_{k,k_{2},k_{1}}^{\alpha(t)}}\right)  }E_{k_{1}}^{T\left(  +\right)  }%
E_{k_{2}}^{T\left(  -\right)  }\delta(k-k_{1}-k_{2})dk_{1}dk_{2};
\label{eq3.3}%
\end{equation}
substituting Eq.(\ref{eq3.3}) into Eq.(\ref{eq2.23}) yields%

\begin{equation}
\left(  {k^{2}c^{2}-\omega^{2}\varepsilon_{k}^{t}}\right)  E_{k}^{TS}=4\pi
i\omega\sum\limits_{\alpha}{\int{\tilde{S}_{k,k_{1},k_{2}}^{\alpha\left(
t\right)  }}}E_{k_{1}}^{T\left(  +\right)  }E_{k_{2}}^{T\left(  -\right)
}\delta\left(  {k-k_{1}-k_{2}}\right)  dk_{1}dk_{2}, \label{eq3.4}%
\end{equation}
where%

\begin{align}
\tilde{S}_{k,k_{1},k_{2}}^{\alpha\left(  t\right)  }  &  \equiv\left(
{\mathrm{\mathbf{S}}_{k,k_{1},k_{2}}^{\alpha(t)}+\mathrm{\mathbf{S}}%
_{k,k_{2},k_{1}}^{\alpha(t)}}\right)  \cdot\mathrm{\mathbf{e}}%
_{\mathrm{\mathbf{k}}}^{t^{*}}=\frac14q_{\alpha}^{3}\int{\frac
{\mathrm{\mathbf{e}}_{\mathrm{\mathbf{k}}}^{t\mathrm{\mathbf{*}}}%
\cdot\mathrm{\mathbf{v}}}{\omega-\mathrm{\mathbf{k}}\cdot\mathrm{\mathbf{v}%
}+i\varepsilon}}\times\label{eq3.5}\\
&  \left\{  {\mathrm{\mathbf{\tilde{e}}}}_{\mathrm{\mathbf{k}}%
_{\mathrm{\mathbf{1}}}}^{t}{\cdot\frac\partial{\partial\mathrm{\mathbf{p}}%
}\frac{\mathrm{\mathbf{e}}_{\mathrm{\mathbf{k}}_{2}}^{t}\cdot\frac
\partial{\partial\mathrm{\mathbf{p}}}}{\omega_{2}-\mathrm{\mathbf{k}%
}_{\mathrm{\mathbf{2}}}\cdot\mathrm{\mathbf{v}}+i\varepsilon}%
+\mathrm{\mathbf{\tilde{e}}}_{\mathrm{\mathbf{k}}_{2}}^{t}\cdot\frac
\partial{\partial\mathrm{\mathbf{p}}}\frac{\mathrm{\mathbf{e}}%
_{\mathrm{\mathbf{k}}_{\mathrm{\mathbf{1}}}}^{t}\cdot\frac\partial
{\partial\mathrm{\mathbf{p}}}}{\omega_{1}-\mathrm{\mathbf{k}}%
_{\mathrm{\mathbf{1}}}\cdot\mathrm{\mathbf{v}}+i\varepsilon}}\right\}
\frac{f_{\alpha}^{R}d\mathrm{\mathbf{p}}}{\left(  {2\pi}\right)  ^{3}%
}.\nonumber
\end{align}
with%

\begin{equation}
\mathrm{\mathbf{\tilde{e}}}_{\mathrm{\mathbf{k}}}^{t}\equiv(1-\frac
{\mathrm{\mathbf{k}}\cdot\mathrm{\mathbf{v}}}\omega)\mathrm{\mathbf{e}%
}_{\mathrm{\mathbf{k}}}^{t}+(\frac{\mathrm{\mathbf{v}}\cdot\mathrm{\mathbf{e}%
}_{\mathrm{\mathbf{k}}}^{t}}\omega)\mathrm{\mathbf{k}}; \label{eq3.6}%
\end{equation}
and the second order distribution function is%

\begin{equation}
f_{\alpha,k}^{\left(  2\right)  }=\int{\Sigma_{k,k_{1},k_{2}}^{\alpha}%
}E_{k_{1}}^{T\left(  +\right)  }E_{k_{2}}^{T\left(  -\right)  }\delta
(k-k_{1}-k_{2})dk_{1}dk_{2}, \label{eq3.7}%
\end{equation}
where%

\begin{equation}
\Sigma_{k,k_{1},k_{2}}^{\alpha}=-q_{\alpha}^{2}\frac{1}{\omega
-\mathrm{\mathbf{k}}\cdot\mathrm{\mathbf{v}}+i\varepsilon}\left\{
{\mathrm{\mathbf{\tilde{e}}}_{\mathrm{\mathbf{k}}_{1}}^{t}\cdot\frac{\partial
}{\partial\mathrm{\mathbf{p}}}\frac{\mathrm{\mathbf{e}}_{\mathrm{\mathbf{k}%
}_{2}}^{t}\cdot\frac{\partial}{\partial\mathrm{\mathbf{p}}}}{\omega
_{2}-\mathrm{\mathbf{k}}_{\mathrm{\mathbf{2}}}\cdot\mathrm{\mathbf{v}%
}+i\varepsilon}+\mathrm{\mathbf{\tilde{e}}}_{\mathrm{\mathbf{k}}_{2}}^{t}%
\cdot\frac{\partial}{\partial\mathrm{\mathbf{p}}}\frac{\mathrm{\mathbf{e}%
}_{\mathrm{\mathbf{k}}_{\mathrm{\mathbf{1}}}}^{t}\cdot\frac{\partial}%
{\partial\mathrm{\mathbf{p}}}}{\omega_{1}-\mathrm{\mathbf{k}}%
_{\mathrm{\mathbf{1}}}\cdot\mathrm{\mathbf{v}}+i\varepsilon}}\right\}
f_{\alpha}^{R}. \label{eq3.8}%
\end{equation}

\section{HIGH-FREQUENCY TRANSVERSE FIELD EQUATION }

If $E_{k}^{T}$ in the left-hand side of Eq.(\ref{eq2.23}) is high-frequency
field, $E_{k}^{T}=E_{k}^{T(+)}$, the quadratic terms in the Eq.(\ref{eq2.19})
should be the product of high-frequency and low-frequency fields:. $E_{k_{1}%
}^{T}E_{k_{2}}^{T}=E_{k_{1}}^{Th}E_{k_{2}}^{TS}+E_{k_{1}}^{TS}E_{k_{2}}^{Th}$.
The three-field product included in the current $\mathrm{\mathbf{j}}%
_{k}^{\left(  3\right)  }$ can be expressed in terms of high-frequency fields
and cubic a mixed product of high-frequency and low-frequency fields. Using
Eq.(\ref{eq3.4}), then, in fact, this mixed term is the product of four
high-frequency fields, which is higher order . Then $E_{k_{1}}^{T}E_{k_{2}%
}^{T}E_{k_{3}}^{T}\approx E_{k_{1}}^{Th}E_{k_{2}}^{Th}E_{k_{3}}^{Th}$. Due to
the factor $\left[  {\left(  {\omega-\omega_{1}}\right)  -\left(
{\mathrm{\mathbf{k}}-\mathrm{\mathbf{k}}_{1}}\right)  \cdot\mathrm{\mathbf{v}%
}+i\varepsilon}\right]  ^{-1}$ in the Eq.(\ref{eq2.22}), its contribution to
the $\mathrm{\mathbf{j}}_{k}^{\left(  3\right)  }$ is important if $E_{k_{1}%
}^{Th}$ is the positive high-frequency fields. As a result%

\[
E_{k_{1}}^{T}E_{k_{2}}^{T}E_{k_{3}}^{T}\approx E_{k_{1}}^{T\left(  +\right)  }
\left[  {E_{k_{2}}^{T\left(  +\right)  }E_{k_{3}}^{T\left(  -\right)
}+E_{k_{2}}^{T\left(  -\right)  }E_{k_{3}}^{T\left(  +\right)  }}\right]  .
\]
Therefore, we gets similarly the high-frequency field equation as follow%

\begin{equation}%
\begin{array}
[c]{l}%
\left(  {k^{2}c^{2}-\omega^{2}\varepsilon_{k}^{t}}\right)  E_{k}^{T\left(
+\right)  }=4\pi i\omega\left[  {\sum\limits_{\alpha}{\int}}\overset{\approx
}{S}{{{_{k,k_{1},k_{2}}^{\alpha\left(  t\right)  }}}E_{k_{1}}^{T\left(
+\right)  }\tilde{E}_{k_{2}}^{TS}\delta\left(  {k-k_{1}-k_{2}}\right)
dk_{1}dk_{2}+}\right. \\
\left.  {\quad\quad\quad\quad\quad\quad\sum\limits_{\alpha}{\int{\tilde
{G}_{k,k_{1},k_{2},k_{3}}^{\alpha(t)}E_{k_{1}}^{T\left(  +\right)  }E_{k_{2}%
}^{T\left(  +\right)  }E_{k_{3}}^{T\left(  -\right)  }\delta\left(
{k-k_{1}-k_{2}-k_{3}}\right)  dk_{1}dk_{2}dk_{3}}}}\right]  ,
\end{array}
\label{eq4.1}%
\end{equation}
where%

\begin{align}
\tilde{G}_{k,k_{1},k_{2},k_{3}}^{\alpha(t)}  &  =\mathbf{e}%
_{\mathrm{\mathbf{k}}}^{t\mathrm{\mathbf{\ast}}}\cdot\left(
\mathrm{\mathbf{G}}_{k,k_{1},k_{2},k_{3}}^{\alpha(t)}+\mathrm{\mathbf{G}%
}_{k,k_{1},k_{3},k_{2}}^{\alpha(t)}\right) \label{eq4.2}\\
&  =-\frac{1}{4}\int{iq_{\alpha}^{4}\frac{(\mathrm{\mathbf{e}}%
_{\mathrm{\mathbf{k}}}^{t\ast}\cdot\mathrm{\mathbf{v}})}{\left(
{\omega-\mathrm{\mathbf{k}}\cdot\mathrm{\mathbf{v}}+i\varepsilon}\right)
}\mathrm{\mathbf{\tilde{e}}}_{\mathrm{\mathbf{k}}_{1}}^{t}\cdot\frac{\partial
}{\partial\mathrm{\mathbf{p}}}\frac{1}{\left[  {\left(  {\omega-\omega_{1}%
}\right)  -\left(  {\mathrm{\mathbf{k}}-\mathrm{\mathbf{k}}_{1}}\right)
\cdot\mathrm{\mathbf{v}}+i\varepsilon}\right]  }\times}\nonumber\\
&  \left\{  {\mathrm{\mathbf{\tilde{e}}}_{\mathrm{\mathbf{k}}_{2}}^{t}%
\cdot\frac{\partial}{\partial\mathrm{\mathbf{p}}}\frac{\mathrm{\mathbf{e}%
}_{\mathrm{\mathbf{k}}_{\mathrm{\mathbf{3}}}}^{t}\cdot\frac{\partial}%
{\partial\mathrm{\mathbf{p}}}}{\omega_{3}-\mathrm{\mathbf{k}}%
_{\mathrm{\mathbf{3}}}\cdot\mathrm{\mathbf{v}}+i\varepsilon}%
+\mathrm{\mathbf{\tilde{e}}}_{\mathrm{\mathbf{k}}_{3}}^{t}\cdot\frac{\partial
}{\partial\mathrm{\mathbf{p}}}\frac{\mathrm{\mathbf{e}}_{\mathrm{\mathbf{k}%
}_{\mathrm{\mathbf{2}}}}^{t}\cdot\frac{\partial}{\partial\mathrm{\mathbf{p}}}%
}{\omega_{2}-\mathrm{\mathbf{k}}_{\mathbf{2}}\cdot\mathrm{\mathbf{v}%
}+i\varepsilon}}\right\}  f_{\alpha}^{R}\frac{d\mathrm{\mathbf{p}}}{(2\pi
)^{3}};\nonumber
\end{align}
and the expression for$\overset{\approx}{S}{{{_{k,k_{1},k_{2}}^{\alpha\left(
t\right)  }}}}$ is the same as $\tilde{S}_{k,k_{1},k_{2}}^{\alpha\left(
t\right)  }$(see Eq.(\ref{eq3.5})), except that $\omega_{2}$ is low-frequency
now. And $\tilde{E}_{k}^{TS}$ is low-frequency fields, but it may different
from the low-frequency fields $\tilde{E}_{k}^{TS}$ in the left-hand of
Eq.(\ref{eq3.4}). According to semiclassical theory, the fusion and decay
interactions in Eqs.(\ref{eq3.4}) and (\ref{eq4.1}) can determine the field
intensity with low-frequency, $N_{k}^{TS}\sim|E_{k}^{TS}|^{2}$, in other
words, they differ by a phase factor $e^{i\phi}$.

The symbol ${\sum\limits_{\alpha}}$ in those expressions above implies adding
the contribution of dark matter($\alpha=2$) and bright matter($\alpha=1$).
From Eqs.(\ref{eq3.5} and \ref{eq4.2}) one can see that $f_{\alpha}%
^{R}d\mathrm{\mathbf{p}}\sim n_{0},q_{\alpha}\sim m_{\alpha},$ $\tilde
{S}^{\alpha}\propto m_{\alpha},G^{\alpha}\propto m_{\alpha}$, namely that the
matrix elements of interaction are proportion to the mass of particles. As the
assumption of the sark matter mass far larger than the bright one[see
Eq.(\ref{eq2.5})], we can neglect the contributions of the bright matter.
Therefore we write $\tilde{S}_{k,k_{1},k_{2}}^{\alpha\left(  t\right)
},\overset{\approx}{S}{{{_{k,k_{1},k_{2}}^{\alpha\left(  t\right)  }}}}$and
$\tilde{G}_{k,k_{1,}k_{2,}k_{3}}^{\alpha(t)}$ as $\tilde{S}_{k,k_{1},k_{2}%
}^{2\left(  t\right)  },\overset{\approx}{S}{{{_{k,k_{1},k_{2}}^{2\left(
t\right)  }}}}$and $\tilde{G}_{k,k_{1,}k_{2,}k_{3}}^{2(t)}$.

In order to get the field equation in spectrum space for nonlinear interaction
up to the third order, we must estimate in detail the integral value of matrix
elements$\tilde{S}_{k,k_{1},k_{2}}^{2\left(  t\right)  }$and $\tilde
{G}_{k,k_{1,}k_{2,}k_{3}}^{2(t)}$ in Eq.(\ref{eq4.1}). Integrating
Eq.(\ref{eq4.2}) by parts and by using of Eq.(\ref{eq2.38}) and (\ref{eq3.8}),
we get%

\begin{equation}
4\pi i\omega\tilde{G}_{k,k_{1,}k_{2,}k_{3}}^{2(t)}\approx4\omega_{p2}%
^{2}\mathrm{\mathbf{\tilde{e}}}_{\mathrm{\mathbf{k}}}^{t\mathrm{\mathbf{\ast}%
}}\cdot\mathrm{\mathbf{\tilde{e}}}_{\mathrm{\mathbf{k}}_{1}}^{t}\frac{1}%
{n_{0}}\int\Sigma_{k-k_{1},k_{2},k_{3}}^{\alpha=2}\frac{d\mathrm{\mathbf{p}}%
}{(2\pi)^{3}}; \label{eq4.3}%
\end{equation}
and for $\overset{\approx}{S}{{{_{k,k_{1},k_{2}}^{\alpha\left(  t\right)  }}}%
}$\ , in which $\omega$ and $\omega_{1}$ are high-frequency and $\omega_{2}$
is low-frequency, after integrating by parts, it reduces to%

\begin{equation}
\overset{\approx}{S}{{{_{k,k_{1},k_{2}}^{2\left(  t\right)  }}}}\approx
\frac{1}{4}\frac{q_{2}^{3}n_{0}}{\omega\omega_{1}\omega_{2}m_{2}^{2}%
}\mathrm{\mathbf{e}}_{\mathrm{\mathbf{k}}}^{t\mathrm{\mathbf{\ast}}}%
\cdot\left[  {\mathrm{\mathbf{e}}_{\mathrm{\mathbf{k}}_{\mathrm{\mathbf{1}}}%
}^{t}\times\left(  {\mathrm{\mathbf{k}}_{\mathrm{\mathbf{2}}}\times
\mathrm{\mathbf{e}}_{\mathrm{\mathbf{k}}_{\mathrm{\mathbf{2}}}}^{t}}\right)
}\right]  . \label{eq4.4}%
\end{equation}
Then first integral of Eq. (\ref{eq4.1}) becomes%

\begin{align}
&  4\pi i\omega\int\overset{\approx}{S}{_{k,k_{1},k_{2}}^{2\left(  t\right)
}E_{k_{1}}^{T(+)}E_{k_{2}}^{TS}\delta\left(  {k-k_{1}-k_{2}}\right)
dk_{1}dk_{2}}\label{eq4.5}\\
&  =4i\frac{q_{2}\omega_{p2}}{m_{2}c}\mathrm{\mathbf{e}}_{\mathrm{\mathbf{k}}%
}^{t\mathrm{\mathbf{\ast}}}\cdot\int{\left[  {\mathrm{\mathbf{E}}_{k_{1}%
}^{T(+)}\times\mathrm{\mathbf{B}}_{k_{2}}^{S}}\right]  }\delta\left(
{k-k_{1}-k_{2}}\right)  dk_{1}dk_{2},\nonumber
\end{align}
where%

\begin{equation}
\mathrm{\mathbf{B}}_{k_{2}}^{s}=\frac{\mathrm{\mathbf{k}}_{2}c}{\omega_{2}%
}\times\mathrm{\mathbf{\tilde{E}}}_{k_{2}}^{TS} \label{eq4.6}%
\end{equation}
is low-frequency magnetic filed, which are produced by the fields with
positive and negative high-frequencies. And using Eq.(\ref{eq3.7}), the second
integral of Eq. (\ref{eq4.1}) becomes%

\begin{equation}
4\omega_{p2}^{2}\int{\mathrm{\mathbf{e}}_{\mathrm{\mathbf{k}}}^{t*}%
\cdot(\mathrm{\mathbf{E}}_{k_{1}}^{T(+)}\frac{n_{k-k_{1}}^{(2)}}{n_{0}}%
)dk_{1}}, \label{eq4.7}%
\end{equation}
where $n_{k^{\prime}}^{(2)}$is the second order of perturbed density,%

\begin{equation}
n_{k^{\prime}}^{(2)}=\int{f_{2,k^{\prime}}^{(2)}\frac{d\mathrm{\mathbf{p}}%
}{\left(  {2\pi}\right)  ^{3}}\quad}. \label{eq4.8}%
\end{equation}
Therefore Eq.(\ref{eq4.1}) is reduced to%

\begin{equation}
\left(  {k^{2}c^{2}-\omega^{2}\varepsilon_{k}^{t}}\right)  \mathrm{\mathbf{E}%
}_{k}^{T(+)}=4\omega_{p2}^{2}\int{(\mathrm{\mathbf{E}}_{k_{1}}^{T(+)}%
\frac{n_{k-k_{1}}^{(2)}}{n_{0}})dk_{1}+4\frac{iq_{2}}{m_{2}c}\omega_{p2}%
\int{\left(  {\mathrm{\mathbf{E}}_{k_{1}}^{T(+)}\times\mathrm{\mathbf{B}%
}_{k-k_{1}}^{S}}\right)  }dk_{1}}. \label{eq4.9}%
\end{equation}
As a result, we obtain from Eq.(\ref{eq4.9})%

\begin{align}
&  \frac{2i}{\omega_{p2}}\frac{\partial\mathrm{\mathbf{E}}(\mathrm{\mathbf{r}%
},t)}{\partial t}-\frac{c^{2}}{\omega_{p2}^{2}}\nabla\times\nabla
\times\mathrm{\mathbf{E}}(\mathrm{\mathbf{r}},t)\label{eq4.10}\\
&  =-2\mathrm{\mathbf{E}}(\mathrm{\mathbf{r}},t)+4\frac{n^{(2)}%
(\mathrm{\mathbf{r}},t)}{n_{0}}\mathrm{\mathbf{E}}(\mathrm{\mathbf{r}%
},t)+4\frac{iq_{2}}{m_{2}c\omega_{p2}}\mathrm{\mathbf{E}}(\mathrm{\mathbf{r}%
},t)\times\mathrm{\mathbf{B}}^{S}(\mathrm{\mathbf{r}},t)\quad,\nonumber
\end{align}
here $\mathrm{\mathbf{E}}(\mathrm{\mathbf{r}},t)$ is the envelope for the
high-frequency fields,%

\begin{equation}
\mathrm{\mathbf{E}}\left(  {\mathrm{\mathbf{r}},t}\right)  e^{-i\omega_{p2}%
t}=\int{\mathrm{\mathbf{E}}_{k}^{T(+)}e^{-i\omega t+i\mathrm{\mathbf{k}}%
\cdot\mathrm{\mathbf{r}}}dk}, \label{eq4.11}%
\end{equation}
and because of slow change in $\mathrm{\mathbf{E}}(\mathrm{\mathbf{r}},t)$, we
have neglected the term $(\frac{\partial^{2}}{\partial t^{2}}%
)\mathrm{\mathbf{E}}(\mathrm{\mathbf{r}},t)$.

\section{PERTURBED DENSITIES }

It is possible that two transverse fields with high frequencies can produce a
longitudinal low-frequency field. In this case, the density perturbation of
the first order can be exited. Then according to Eq. (\ref{eq3.4}), one has%

\begin{equation}
\varepsilon_{k}^{l}E_{k}^{TS(l)}\;=-\frac{4\pi i}\omega\sum\limits_{\alpha
}{\int{\tilde{S}_{k,k_{1},k_{2}}^{2\left(  l\right)  }}}E_{k_{1}}^{T\left(
+\right)  }E_{k_{2}}^{T\left(  -\right)  }\delta\left(  {k-k_{1}-k_{2}%
}\right)  dk_{1}dk_{2}\quad, \label{eq5.1}%
\end{equation}
where the coupling matrix $\tilde{S}_{k,k_{1},k_{2}}^{2(l)}$ is the same as
$\tilde{S}_{k,k_{1},k_{2}}^{2(t)}$ when $\mathrm{\mathbf{e}}%
_{\mathrm{\mathbf{k}}}^{t*}\rightarrow\mathrm{\mathbf{k/}}k$. the density
perturbation produced by the longitudinal low-frequency wave is%

\begin{equation}
n_{k}^{(1)}=\int{f_{k}^{T(1)}\frac{d\mathrm{\mathbf{p}}}{\left(  {2\pi
}\right)  ^{3}}}=\frac{q_{2}}{i}E_{k}^{TS(l)}\int{\frac{1}{k}\frac
{\mathrm{\mathbf{k}}\cdot\frac{\partial f_{2}^{R}}{\partial\mathrm{\mathbf{p}%
}}}{\omega-\mathrm{\mathbf{k}}\cdot\mathrm{\mathbf{v}}+i\varepsilon}%
\frac{d\mathrm{\mathbf{p}}}{\left(  {2\pi}\right)  ^{3}}}. \label{eq5.2}%
\end{equation}
Taking account of the integral expression of $\varepsilon_{k}^{2(l)}$[see
Eq.(\ref{eq2.25})], we obtain%

\begin{equation}
n_{k}^{(1)}=-\frac{q_{2}}i\frac k{\pi q_{2}^{2}}\left(  {\varepsilon
_{k}^{2(l)}-1}\right)  E_{k}^{TS(l)}. \label{eq5.3}%
\end{equation}
Using $\left(  {\varepsilon_{k}^{2(l)}-1}\right)  =\varepsilon_{k}%
^{l}-\varepsilon_{k}^{1(l)}$, and $\varepsilon_{k}^{l}E_{k}^{TS(l)}\;$, as the
first-order approximation, is zero[see Eq.(\ref{eq5.1})], then%

\begin{equation}
n_{k}^{(1)}=\frac{q_{2}}i\frac k{\pi q_{2}^{2}}\varepsilon_{k}^{1(l)}%
E_{k}^{TS(l)}. \label{eq5.4}%
\end{equation}
Similarly, one estimate the matrix as follow:%

\begin{equation}
\tilde{S}_{k,k_{1},k_{2}}^{2\left(  l\right)  }\approx-\frac{k\omega}{\pi
m_{2}\omega_{p2}^{2}}q_{2}\left(  {\varepsilon_{k}^{2(l)}-1}\right)  \left(
\mathbf{e}_{\mathbf{k}_{2}}^{t}\cdot\mathbf{e}_{\mathbf{k}_{1}}^{t}\right)  .
\label{eq5.5}%
\end{equation}
So the total perturbed density produced by two transverse fields with positive
and negative high-frequencies, is%

\begin{align}
{n}_{k}^{\prime}  &  =n_{k}^{(1)}+n_{k}^{(2)}\label{eq5.6}\\
&  =-4\frac{k^{2}}{\pi m_{2}\omega_{p2}^{2}}\frac{-\varepsilon_{k}%
^{1(l)}+\varepsilon_{k}^{l}}{\varepsilon_{k}^{l}}\left(  {\varepsilon
_{k}^{2(l)}-1}\right)  \int{\mathrm{\mathbf{E}}_{k_{1}}^{T(+)}\cdot
\mathrm{\mathbf{E}}_{k_{2}}^{T(-)}\delta\left(  {k-k_{1}-k_{2}}\right)
dk_{1}dk_{2}.}\nonumber
\end{align}

Then, in the coordinate representation, Eq.(\ref{eq5.6}) becomes%

\begin{equation}
n^{\prime}(\mathrm{\mathbf{r}},t)=4\frac{\left\vert {\mathrm{\mathbf{E}%
}(\mathrm{\mathbf{r}},t)}\right\vert ^{2}}{\pi T_{2}}. \label{eq5.7}%
\end{equation}
In the case of taking account of longitudinal low-frequency fields, the first
term to the right-hand side of Eq. (\ref{eq4.9}) will be added to the coupling
term of longitudinal fields,%

\begin{equation}
4\pi i\omega\int\overset{\approx}{S}{{{_{k,k_{1},k_{2}}^{2\left(  l\right)  }%
}}}E_{k_{1}}^{T\left(  +\right)  }E_{k_{2}}^{TS\left(  l\right)  }%
\delta\left(  {k-k_{1}-k_{2}}\right)  dk_{1}dk_{2}\quad,\label{eq5.8}%
\end{equation}
where the coupling matrix $\overset{\approx}{S}{{{_{k,k_{1},k_{2}}^{2\left(
l\right)  }}}}$is similar to $\overset{\approx}{S}{{{_{k,k_{1},k_{2}%
}^{2\left(  l\right)  }}}}(e_{\mathrm{\mathbf{k}}_{2}}^{t}\rightarrow
\frac{\mathrm{\mathbf{k}}_{2}}{k_{2}})$ in Eq. (\ref{eq4.5}), its estimated
value is
\begin{equation}
\overset{\approx}{S}{{{_{k,k_{1},k_{2}}^{2\left(  l\right)  }}}}\approx
-\frac{1}{4}q_{2}^{3}\int{\frac{\mathrm{\mathbf{e}}_{\mathrm{\mathbf{k}}%
}^{t\mathrm{\mathbf{\ast}}}\cdot\mathrm{\mathbf{e}}_{\mathrm{\mathbf{k}%
}_{\mathrm{\mathbf{1}}}}^{t}}{\omega m_{2}}\cdot\frac{\frac{\mathrm{\mathbf{k}%
}_{2}}{k_{2}}\cdot\frac{\partial}{\partial\mathrm{\mathbf{p}}}f_{2}^{R}%
}{\omega_{2}-\mathrm{\mathbf{k}}_{\mathrm{\mathbf{2}}}\cdot\mathrm{\mathbf{v}%
}+i\varepsilon}\cdot\frac{d\mathrm{\mathbf{p}}}{\left(  {2\pi}\right)  ^{3}}}.
\label{eq5.9}%
\end{equation}
And Eq. (\ref{eq5.8}) is reduced to%

\begin{equation}
4\omega_{p2}^{2}\frac1{n_{0}}\mathrm{\mathbf{e}}_{\mathrm{\mathbf{k}}%
}^{t\mathrm{\mathbf{*}}}\cdot\int{\mathrm{\mathbf{E}}_{k_{\mathrm{\mathbf{1}}%
}}^{T(+)}}\left[  {\int{f_{k-k_{1}}^{T(1)}}\frac{d\mathrm{\mathbf{p}}}{\left(
{2\pi}\right)  ^{3}}}\right]  dk_{1}=4\omega_{p2}^{2}\mathrm{\mathbf{e}%
}_{\mathrm{\mathbf{k}}}^{t\mathrm{\mathbf{*}}}\cdot\int{\mathrm{\mathbf{E}%
}_{k_{\mathrm{\mathbf{1}}}}^{T(+)}\frac{n_{k-k_{1}}^{(1)}}{n_{0}}}dk_{1}.
\label{eq5.10}%
\end{equation}
Taking this term into account, one gets from Eq.(\ref{eq4.9})%

\begin{equation}
\frac{2i}{\omega_{p2}}\frac{\partial\mathrm{\mathbf{E}}(\mathrm{\mathbf{r}%
},t)}{\partial t}-\frac{c^{2}}{\omega_{p2}^{2}}\nabla\times\nabla
\times\mathrm{\mathbf{E}}(\mathrm{\mathbf{r}},t)+2\mathrm{\mathbf{E}%
}(\mathrm{\mathbf{r}},t) \label{eq5.11}%
\end{equation}

\[
-4\frac{{n}^{\prime}(\mathrm{\mathbf{r}},t)}{n_{0}}\mathrm{\mathbf{E}%
}(\mathrm{\mathbf{r}},t)-4\frac{iq_{2}}{m_{2}c\omega_{p2}}\mathrm{\mathbf{E}%
}(\mathrm{\mathbf{r}},t)\times\mathrm{\mathbf{B}}^{S}(\mathrm{\mathbf{r}%
},t)=0.
\]

\[
\]

\section{NON-LINEAR COUPLING FIELDS EQUATION }

Now we study the low-frequency field equation (\ref{eq3.4}). Differentiating
with respect to section $\mathrm{\mathbf{v}}$ in braces in Eq.(\ref{eq3.5})
and taking account to delta function in Eq.(\ref{eq3.4}), yields%

\begin{equation}
\tilde{S}_{k,k_{1},k_{2}}^{2\left(  t\right)  }\approx-\frac14\frac{q_{2}^{3}%
}{m_{2}^{2}\omega_{p2}^{2}}\frac{n_{0}}\omega\left(  {\frac\omega{\sqrt
{2}kv_{T2}}}\right)  \mathrm{\mathbf{e}}_{\mathrm{\mathbf{k}}}^{t*}\left[
{\mathrm{\mathbf{e}}_{\mathrm{\mathbf{k}}_{1}}^{t}\left(  {\mathrm{\mathbf{k}%
}\cdot\mathrm{\mathbf{e}}_{\mathrm{\mathbf{k}}_{2}}^{t}}\right)
-\mathrm{\mathbf{e}}_{\mathrm{\mathbf{k}}_{2}}^{t}\left(  {\mathrm{\mathbf{k}%
}\cdot\mathrm{\mathbf{e}}_{\mathrm{\mathbf{k}}_{1}}^{t}}\right)  }\right]  .
\label{eq6.1}%
\end{equation}
Then Eq. (\ref{eq3.4}) becomes%

\begin{equation}
\left(  {k^{2}c^{2}-\omega^{2}\varepsilon_{k}^{t}}\right)  \mathrm{\mathbf{E}%
}_{k}^{TS}=-\frac{iq_{2}}{m_{2}}Z\left(  {\frac{\omega}{\sqrt{2}kv_{T2}}%
}\right)  \int{\mathrm{\mathbf{k}}\times\left(  {\mathrm{\mathbf{E}}_{k_{1}%
}^{T(+)}\times\mathrm{\mathbf{E}}_{k_{2}}^{T(-)}}\right)  }\delta\left(
{k-k_{1}-k_{2}}\right)  dk_{1}dk_{2}. \label{eq6.2}%
\end{equation}
According to Eq.(\ref{eq4.6}),%

\begin{equation}
e^{-i\phi}\mathrm{\mathbf{B}}_{k}^{S}=\frac c\omega\mathrm{\mathbf{k}}%
\times\mathrm{\mathbf{E}}_{k}^{TS}, \label{eq6.3}%
\end{equation}
we have%

\begin{align}
\left(  {k^{2}c^{2}-\omega^{2}\varepsilon_{k}^{t}}\right)  \mathrm{\mathbf{B}%
}_{k}^{S}e^{-i\phi}  &  =-\frac{iq_{2}c}{m_{2}\omega}Z\left(  {\frac{\omega
}{\sqrt{2}kv_{T2}}}\right)  \times\label{eq6.4}\\
&  \mathrm{\mathbf{k}}\times\left[  {\mathrm{\mathbf{k}}\times\int{\left(
{\mathrm{\mathbf{E}}_{k_{1}}^{T(+)}\times\mathrm{\mathbf{E}}_{k_{2}}^{T(-)}%
}\right)  }\delta\left(  {k-k_{1}-k_{2}}\right)  dk_{1}dk_{2}}\right]
.\nonumber
\end{align}
For very low frequency fields, $v_{T1}\gg\omega/k\ll v_{T2}$, one has%

\[
Z\left(  {\frac{\omega}{\sqrt2 kv_{T\alpha} }} \right)  \approx\frac
{\omega^{2}}{k^{2}v_{T\alpha}^{2} } - i\sqrt{\frac{\pi}{2}} \frac{\omega
}{kv_{T\alpha} } \approx- i\sqrt{\frac{\pi}{2}} \frac{\omega}{kv_{T\alpha} },
\]

\begin{equation}
\varepsilon_{k}^{t}\approx1+\frac{\omega_{p1}^{2}}{k^{2}v_{T1}^{2}}%
+\frac{\omega_{p2}^{2}}{k^{2}v_{T2}^{2}}-i\sqrt{\frac\pi2}\frac{\omega_{p2}%
}\omega\frac\omega{kv_{T\alpha}}\approx-i\sqrt{\frac\pi2}\frac{\omega_{p2}%
^{2}}{\omega kv_{T2}}; \label{eq6.5a}%
\end{equation}
and the following condition obviously is satisfied%

\begin{equation}
\omega_{p2}\gg\left(  {\frac{kv_{T2}}\omega}\right)  kc. \label{eq6.5b}%
\end{equation}
Then Eq. (\ref{eq6.4}) become%

\begin{equation}
\omega\mathrm{\mathbf{B}}_{k}^{S}e^{-i\phi}=\frac{iq_{2}c}{m_{2}\omega
_{p2}^{2}}\mathrm{\mathbf{k}}\times\left[  {\mathrm{\mathbf{k}}\times
\int{\left(  {\mathrm{\mathbf{E}}_{k_{1}}^{T(+)}\times\mathrm{\mathbf{E}%
}_{k_{2}}^{T(-)}}\right)  }\delta\left(  {k-k_{1}-k_{2}}\right)  dk_{1}dk_{2}%
}\right]  . \label{eq6.6}%
\end{equation}
Hence the coordinate representation of Eq.(\ref{eq6.6}) is reduced to the
following equation%

\begin{equation}
\frac\partial{\partial t}\mathrm{\mathbf{B}}^{S}(\mathrm{\mathbf{r}}%
,t)=i\frac{q_{2}c}{m_{2}\omega_{p2}^{2}}\nabla\times\nabla\times\left[
{{\mathrm{\mathbf{E}}(\mathrm{\mathbf{r}},t)\times\mathrm{\mathbf{E}}%
^{*}(\mathrm{\mathbf{r}},t)}}\right]  , \label{eq6.7}%
\end{equation}
where we have chose $\phi=\frac\pi2$ for getting real magnetic fields with
low-frequency. Through the substitutions%

\begin{align}
\xi &  =\frac{2}{3}\sqrt{\mu}\frac{\mathrm{\mathbf{r}}}{v_{T2}/\omega_{p2}%
},\quad\tau=\frac{2}{3}\mu\omega_{p2}t,\quad\mu=\frac{m_{1}}{m_{2}}%
,\quad\alpha=\frac{c^{2}}{3v_{T2}^{2}},\label{eq6.8}\\
\mathrm{\mathbf{E}}(\xi,\tau)  &  =\frac{4\sqrt{3}\mathrm{\mathbf{E}%
}(\mathrm{\mathbf{r}},t)}{\sqrt{\pi\mu n_{0}T_{2}}},\quad\mathrm{\mathbf{B}%
}(\xi,\tau)=\frac{12q_{2}}{4\mu m_{2}c\omega_{p2}}\mathbf{B}^{s}%
(\mathrm{\mathbf{r}},t),\quad n=\frac{3}{4}\frac{n^{\prime}}{n_{0}},\nonumber
\end{align}
we can now write Eqs.(\ref{eq5.7}), (\ref{eq5.11}) and (\ref{eq6.7}) in the form%

\begin{equation}
n(\xi,\tau)=\left|  {\mathrm{\mathbf{E}}(\xi,\tau)}\right|  ^{2}\quad,
\label{eq6.9}%
\end{equation}

\begin{equation}
i\frac{\partial\mathrm{\mathbf{E}}(\xi,\tau)}{\partial\tau}-\alpha\nabla
\times\nabla\times\mathrm{\mathbf{E}}(\xi,\tau)+\frac3{2\mu}\mathrm{\mathbf{E}%
}(\xi,\tau)-n(\xi,\tau)\mathrm{\mathbf{E}}(\xi,\tau)-i\mathrm{\mathbf{E}}%
(\xi,\tau)\times\mathrm{\mathbf{B}}(\xi,\tau)=0, \label{eq6.10}%
\end{equation}

\begin{equation}
\frac\partial{\partial\tau}\mathrm{\mathbf{B}}(\xi,\tau)=i\frac16\nabla
\times\nabla\times\left[  {\mathrm{\mathbf{E}}(\xi,\tau)\times
\mathrm{\mathbf{E}}^{*}(\xi,\tau)}\right]  . \label{eq6.11}%
\end{equation}

Therefore it may be seen that the gravitoelectric and gravitomagnetic fields,
and self-generated gravito-magnetic fields with very low-frequency are
completely determined by the closed Eqs.(\ref{eq6.9})-(\ref{eq6.11}).

\section{NUMERICAL INTEGRAL}

It is well known that the gravitational system can emit energy in quadrupole
radiation. For example, consider a neutron star with mass $M=M_{\odot}$, its
radius, rotation inertia, rotational period and eccentricity are as follows:%

\[
R\sim10km,\quad M\sim M_{\odot},\quad I\sim10^{45},\quad P\sim0.033,\quad
e\sim10^{-4}.
\]
As a result, due to the quadrupole radiation, the energy loss for the system
is\cite{Ostricker}%

\[
L\approx10^{45}e^{2}\sim10^{37}\quad(erg\cdot s^{-1}).
\]
Furthermore, one can estimate the radiation flux at $r=D=r_{i}=10\times
10km=10^{7}$(cm),%

\[
F\sim L/D^{2}\sim L/r_{i}^{2}\sim10^{23}\quad(erg\cdot cm^{-2}\cdot s^{-1}),
\]
where $r_{i}$ is the inner radius of accretion desk around the compact star ,
at which the corresponding parameters are%

\[
T_{0}=T_{1}=T_{2}=10^{7}K,\quad\rho_{1}\equiv n_{0}m_{p}=\rho_{2}/10=4(g\cdot
cm^{-3})\text{.}
\]
Then we get immediately the estimated values of the field in the gravitational wave%

\[
F\sim c\frac{\left\vert {\mathrm{\mathbf{E}}(\mathrm{\mathbf{r}}%
,t)}\right\vert ^{2}}{8\pi},\quad\frac{\left\vert {\mathrm{\mathbf{E}%
}(\mathrm{\mathbf{r}},t)}\right\vert ^{2}}{8\pi}\sim F/c\sim3.3\times
10^{12}(erg\cdot cm^{-3})\text{, }
\]
i.e.%

\[
\frac{\left|  {\mathrm{\mathbf{E}}(\mathrm{\mathbf{r}},t)}\right|  ^{2}}{8\pi
n_{0}T_{0}}=\bar{W}=\frac{3.3\times10^{14}}{1.38\times10^{-16}n_{0}T_{0}%
}=\frac{3.3\times10^{14}}{1.38\times2.4\times10^{16}}=10^{-3};
\]
and%

\[
\left\vert {\mathrm{\mathbf{E}}(\mathrm{\mathbf{r}},t)}\right\vert _{\tau
=0}^{2}=\frac{384}{\mu}\overline{W};
\]
hence%

\[
\left|  {\mathrm{\mathbf{E}}(\xi,\tau)}\right|  _{\tau=0}^{2}=3.84,\quad
\alpha=\frac{3.8\times10^{13}}{T_{2}}=3.8\times10^{6},\quad(\overline{W}%
_{\tau=0}=10^{-3}).
\]

We have solved numerically Eqs.(\ref{eq6.9}), (\ref{eq6.10}) and
(\ref{eq6.11}) in two dimensions with three field components using FFT. The
initial condition with a periodic boundary condition is given as\cite{Li2,Liu}%

\begin{equation}
\left\vert {\mathrm{\mathbf{E}}(\xi,\tau)}\right\vert _{\tau=0}^{2}=E_{0}%
\sin\frac{2\pi y}{y_{0}}\sec h(\frac{x}{L{\ }_{0}})(\mathrm{\mathbf{e}}%
_{x}+\mathrm{\mathbf{e}}_{z})-E_{0}\frac{y_{0}}{2\pi L_{0}}\cos\frac{2\pi
y}{y_{0}}\tanh\frac{x}{L_{0}}\sec h\frac{x}{L_{0}}\mathrm{\mathbf{e}}_{y}
\label{eq7.1}%
\end{equation}
with%

\[
L_{0}=2\times10^{3},\quad y_{0}=5\times10^{6}.
\]

The distribution of intial gravitoelectric field is shown in Fig.1,
the dynamic evolution behavior for gravitoelectric fields and
self-generated gravitomagnetic fields with the very low frequency is
shown in Figs.2-7 and Figs.8-13. Quantites in Figs. 1-13 are dimensionless. The relations to dimensional ones are%

\begin{equation}
r=3\times10^{6}\sqrt{\frac{T_{2}}{\rho_{2}}}x_{Fig}=1.5\times10^{9}x_{Fig},
\label{eq7.2}%
\end{equation}

\begin{equation}
t=1.6\times10^{4}\frac\tau{\sqrt{\rho_{2}}}=2.5\times10^{3}\tau, \label{eq7.3}%
\end{equation}

\begin{equation}
\frac{\left\vert {\mathrm{\mathbf{E}}(\mathrm{\mathbf{r}},t)}\right\vert ^{2}%
}{8\pi}=3.6\times10^{-20}(n_{0}T_{0})\text{ }\left\vert \mathrm{\mathbf{E}%
}\right\vert _{Fig}^{2}=8.6\times10^{11}\left\vert \mathrm{\mathbf{E}%
}\right\vert _{Fig}^{2}, \label{eq7.4}%
\end{equation}

\begin{equation}
B^{S}=5.6\times10^{9}B_{Fig}. \label{eq7.5}%
\end{equation}

Figs.2-7 give the collapse development of gravitoelectric field
 and Figs.8-13 the collapse development of self-generated
gravitomagnetic field with very low frequency. While $\tau=1.65$,
the strengths of the self-generated gravitomagnetic fields  are%

\[
|\mathrm{\mathbf{E}}|_{\max}^{2}=2748.75\quad\text{, }|\mathrm{\mathbf{B}%
}|_{\max}^{2}=8.64\times10^{-9}.
\]
\bigskip
Using Eqs.(\ref{eq7.2}) --(\ref{eq7.5}), while $t=3.3\times10^{3}s$ \ we have%

\[
\quad|\mathrm{\mathbf{E}}(\mathrm{\mathbf{r}},t)|_{\max}^{2}=5.94\times
10^{16}erg/cm^{3}
\]
and%

\[
\quad B^{s}=5.1\times10^{5}(erg/cm^{3})^{\frac{1}{2}}.
\]

When $\tau>1.65$, the collapses rapidly and leads to a very strong field, i.e.
$\bar{W}=\left\vert {\mathrm{\mathbf{E}}(\mathrm{\mathbf{r}},t)}\right\vert
^{2}/8\pi n_{0}T_{0}>1$ and in this case the expansion Eq.(\ref{eq2.7a}) is no
longer valid.
\section{CONCLUSION REMARKS}
From the above study, we arrive at the following conclusions: If
there is very intense gravitational radiation with high frequency
near by a source, the interactions of the wave decay and fusion and
wave-particle develop, and nonlinear (matter) currents with low
frequency are induced by the interactions, leading to increasing of
local matter density and excitation of a very low-frequency
gravitomagnetic field.

It is shown that the dynamic behavior and configuration for the GEM fields,
the perturbed density and the gravitomagnetic field with very low-frequency
are determined by Eqs.(\ref{eq6.9}) -(\ref{eq6.11})

We investigate the numerical solution of Eqs.(\ref{eq6.9}) -(\ref{eq6.11}). It
is shown that the fields, involving the GEM fields, the perturbed density
field and self-generated gravitomagnetic field with very low frequency, may
collapse, so that the gravitational waves are amplified by a factor of
$10^{3}$. In other words, due to self-condensing, a stronger GME fields could
be produced; and they could appear as the gravitational waves with high energy
reaching on Earth. In this case, Weber's results, perhaps, are acceptable.

Meanwhile, the increasing perturbed matter density and
gravitomagnetic field with very low-frequency in the local region,
where the gravitational waves get through, are in favor of the
formation of a new rotating object.//
\bigskip
\begin{acknowledgments}
This work was partly supported by the National Natural Science Foundation of
China and the Natural Science Foundation of Jiangxi Province.
\end{acknowledgments}

*Response author. Email: sqliu@ncu.edu.cn


\begin{figure}
\epsfig{file = 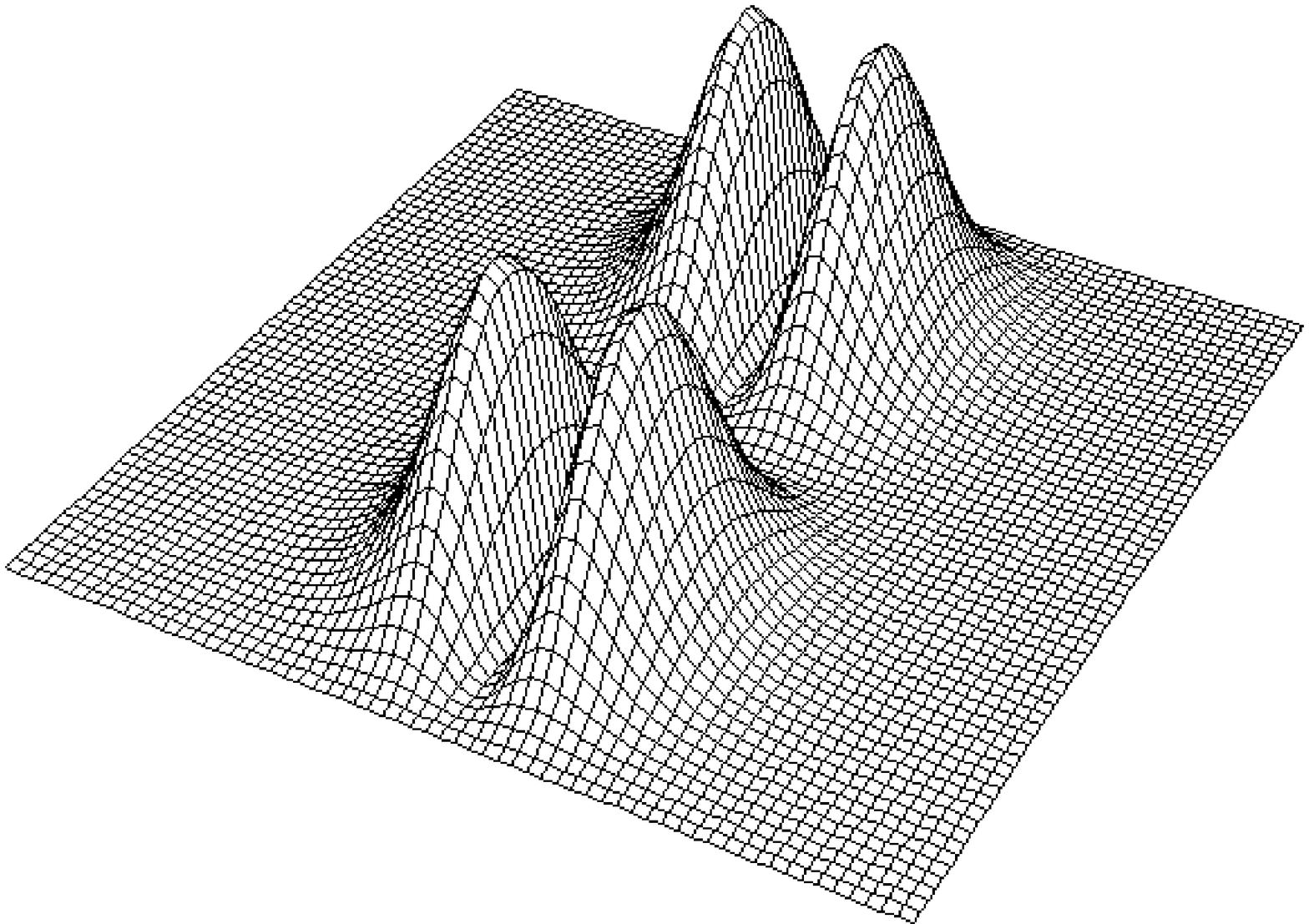, width = 11cm} \caption{Initial gravitoelectric
field distribution:  $(a)\;\tau= 0.00,\left| \mathrm{\mathbf{E}}
\right|  _{\max}^{2} = \left\vert E\right\vert _{x}^{2}+\left\vert
E\right\vert _{y}^{2}+\left\vert E\right\vert _{z}^{2}=3.84$.}
\end{figure}
\bigskip
\begin{figure}
\epsfig{file = 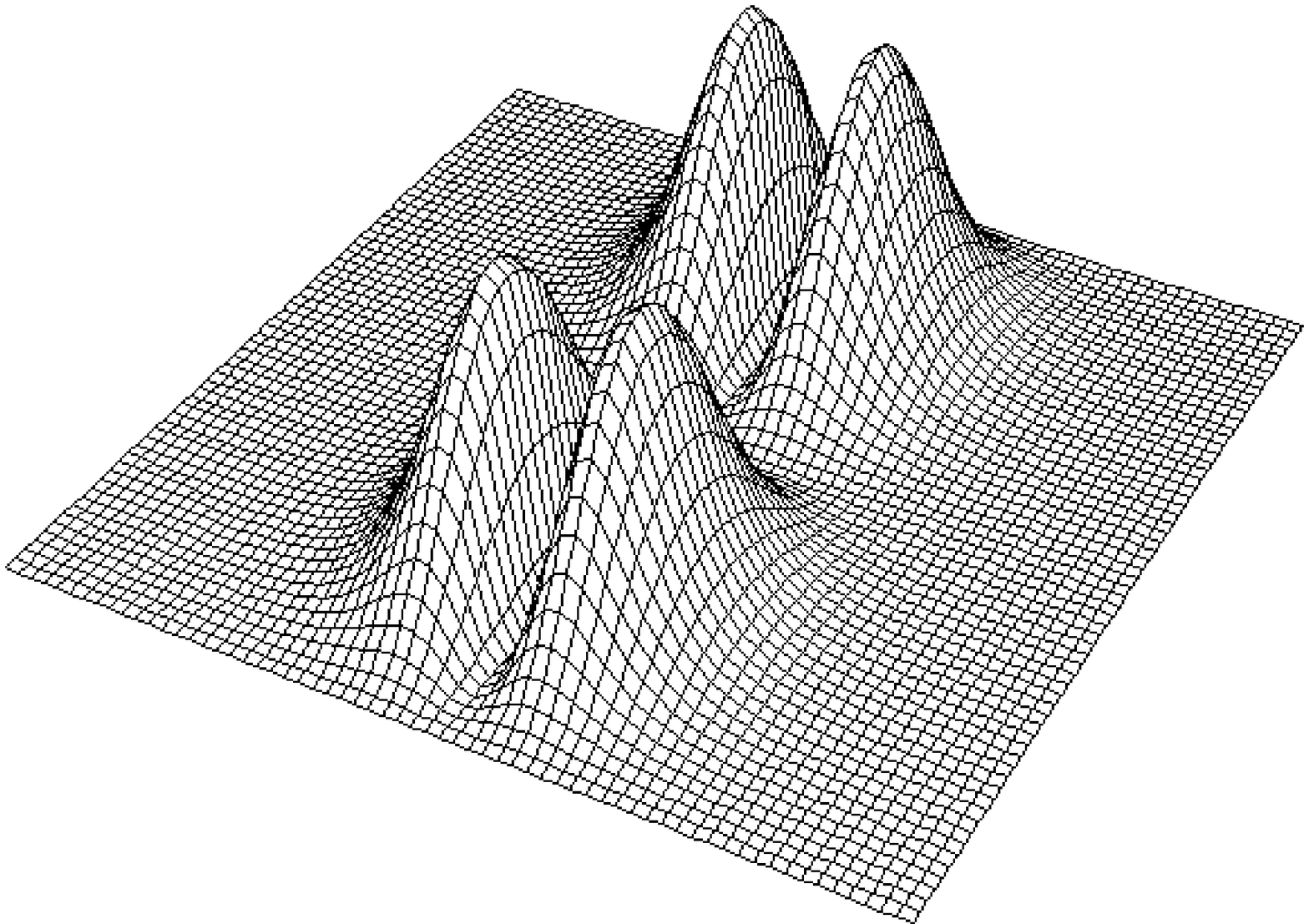, width = 11cm} \caption{$ (a)\;\tau= 0.05,\left|
\mathrm{\mathbf{E}} \right|  _{\max}^{2} = 3.895 $ }
\end{figure}

\begin{figure}
\epsfig{file = 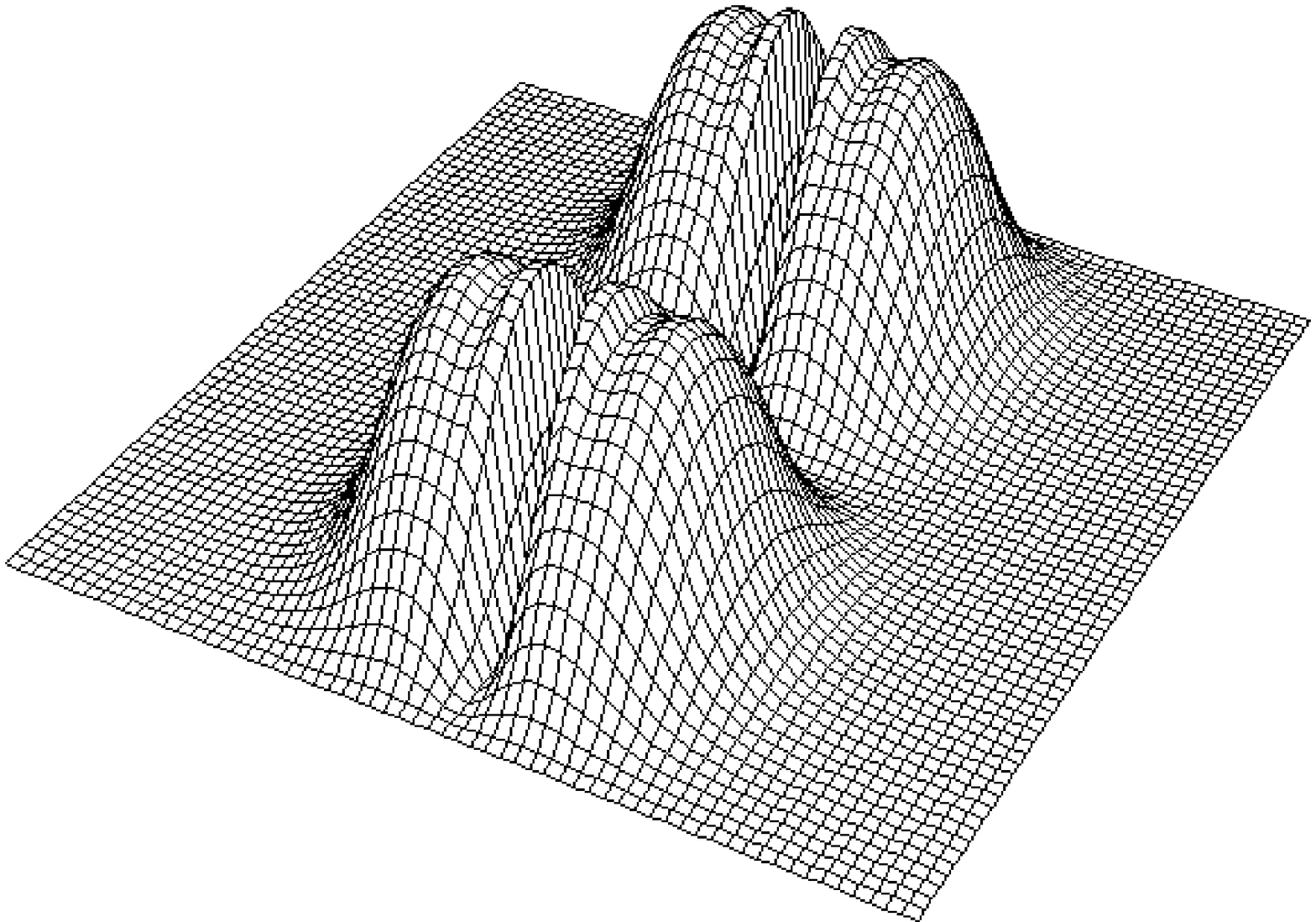, width = 11cm} \caption{$  (b)\;\tau= 0.5,\left|
\mathrm{\mathbf{E}} \right|  _{\max}^{2} = 3.418  $ }
\end{figure}

\begin{figure}
\epsfig{file = 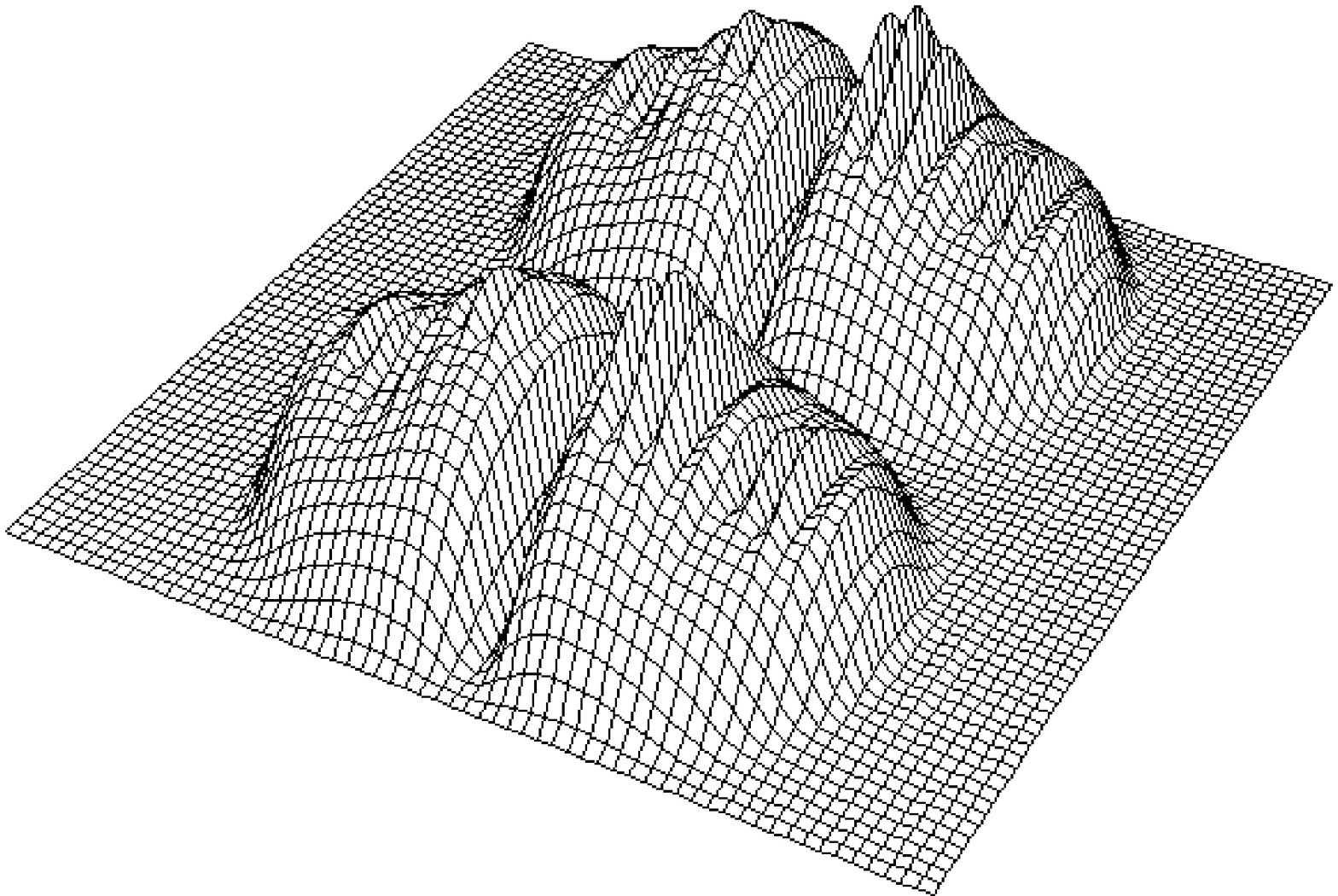, width = 11cm} \caption{$ (c)\;\tau= 1.4,\left|
\mathrm{\mathbf{E}} \right|  _{\max}^{2} = 5.213 $ }
\end{figure}

\begin{figure}
\epsfig{file = 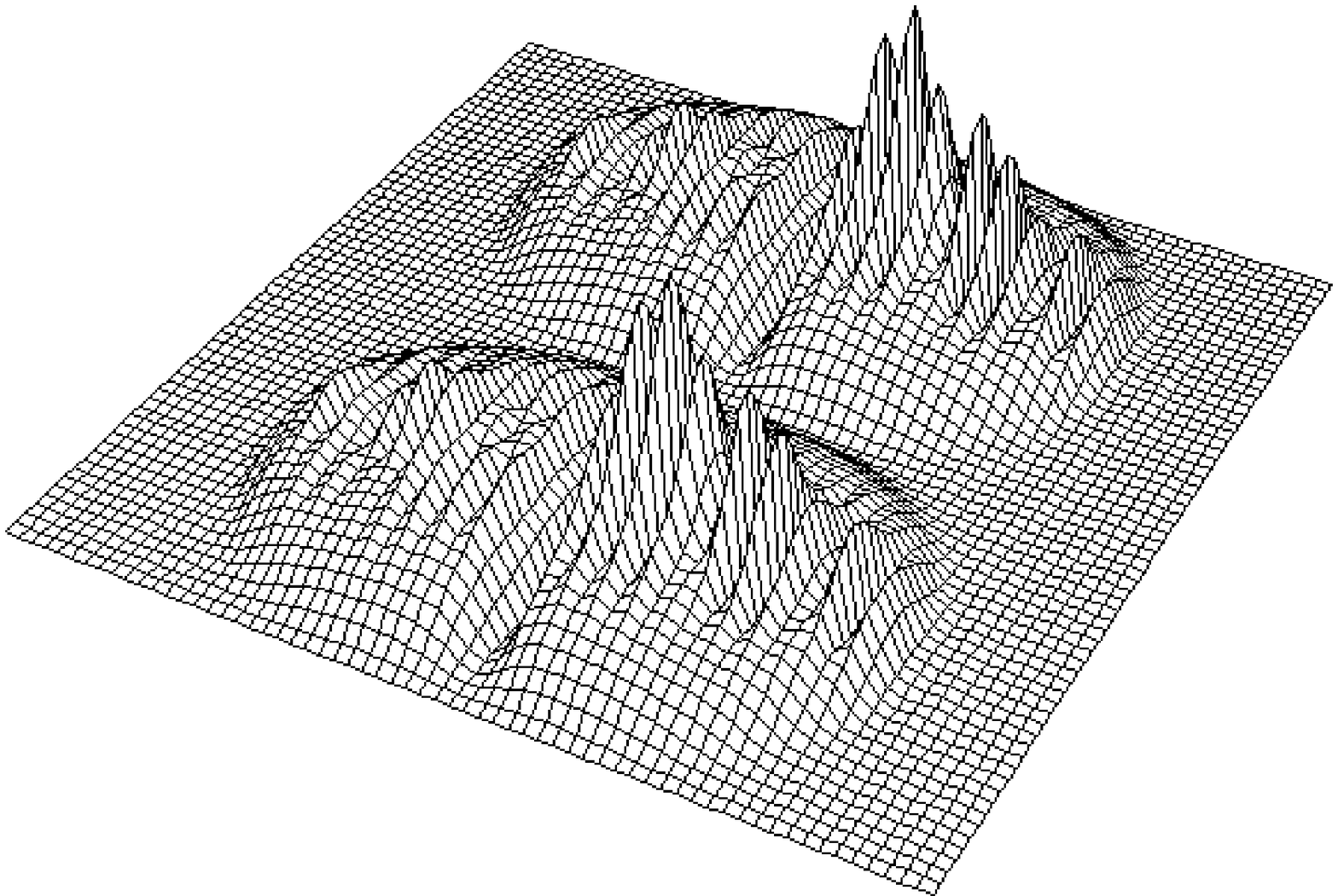, width = 11cm} \caption{$ (d)\;\tau= 1.55,\left|
\mathrm{\mathbf{E}} \right|  _{\max}^{2} = 12.09 $ }
\end{figure}

\begin{figure}
\epsfig{file = 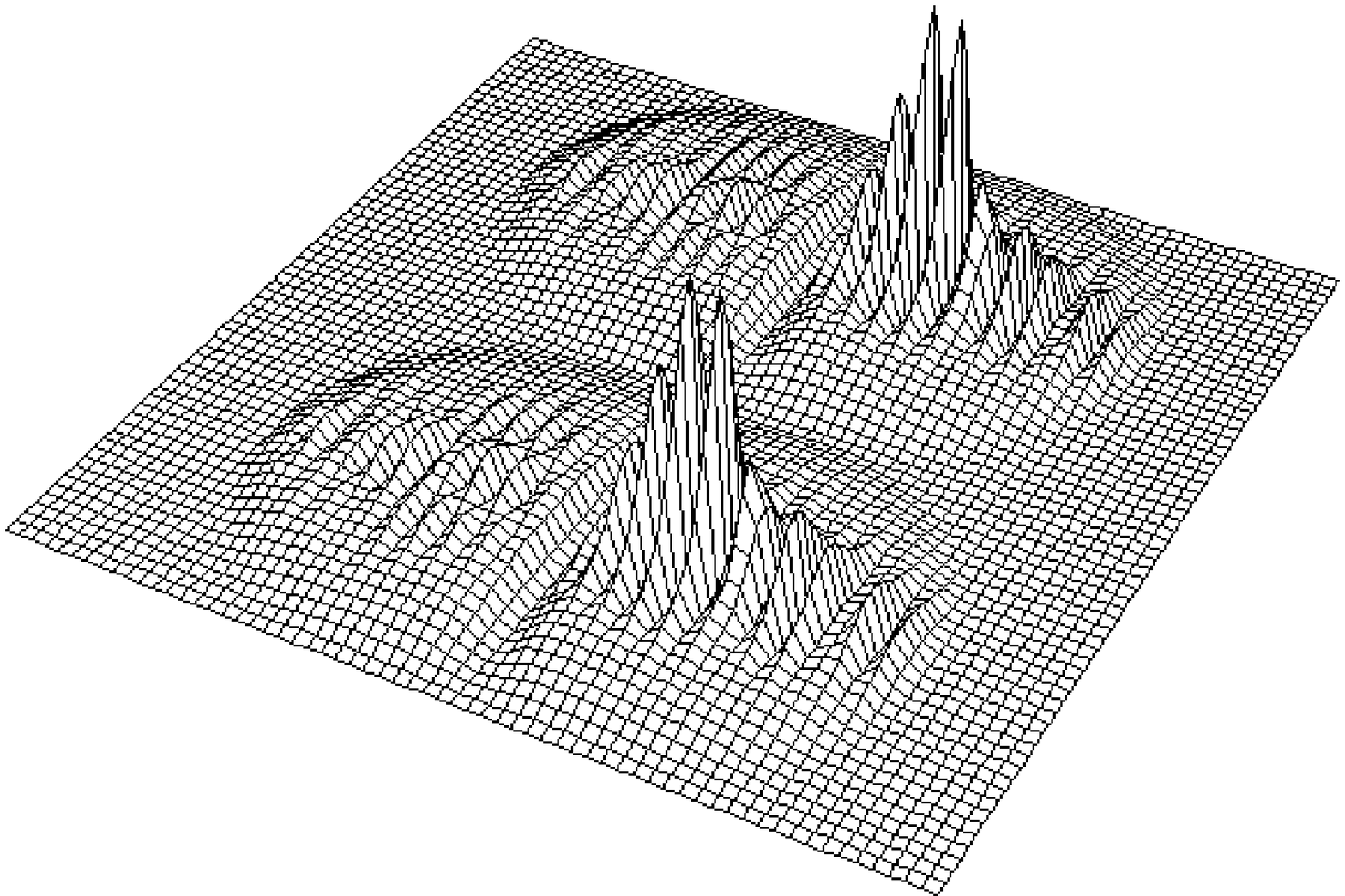, width = 11cm} \caption{$ (e)\;\tau=1.6,\left|
\mathrm{\mathbf{E}}\right|  _{\max}^{2}=30.55$ }
\end{figure}

\begin{figure}
\epsfig{file = 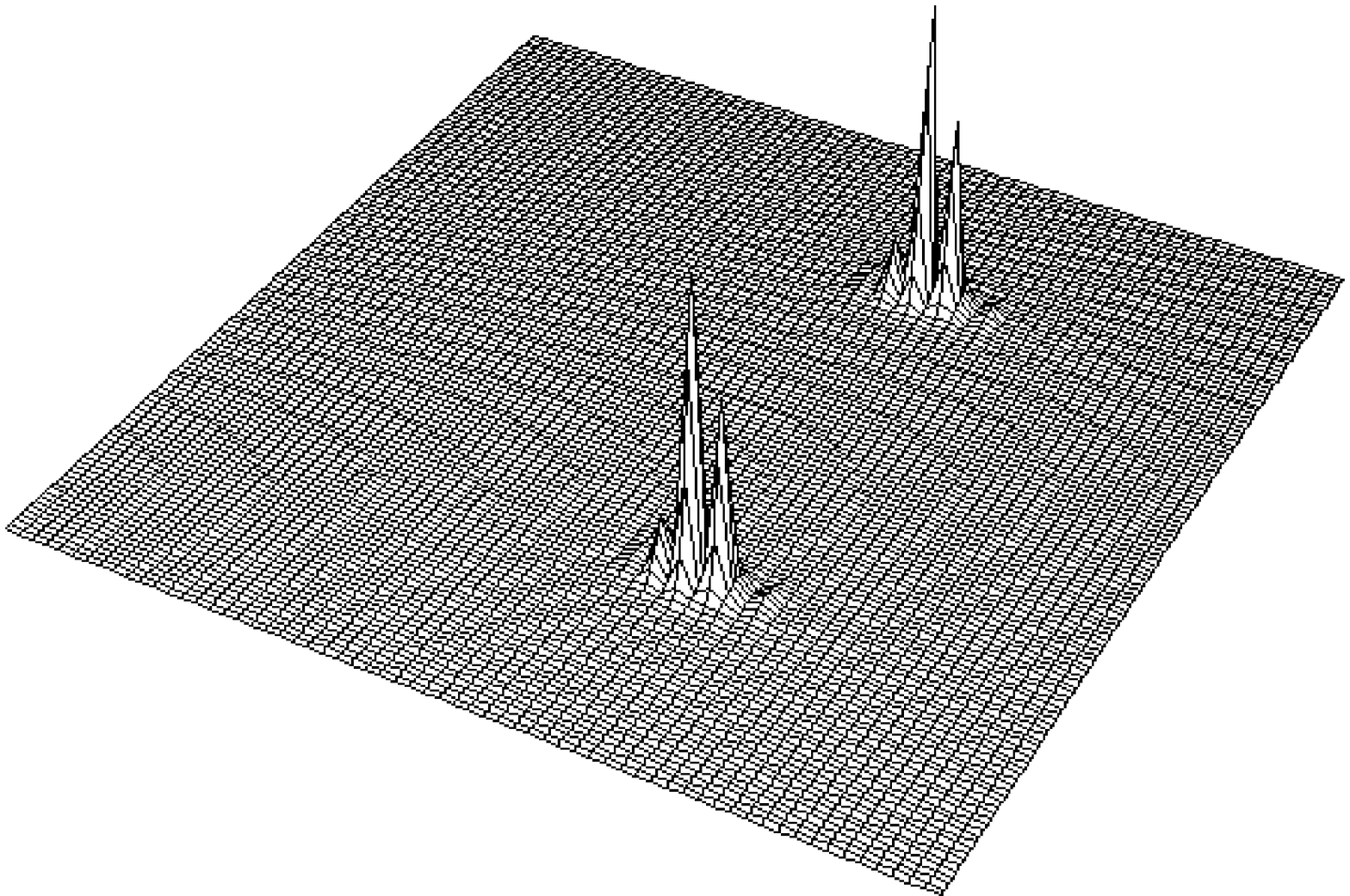, width = 11cm} \caption{$ (f)\;\tau=1.65,\left|
\mathrm{\mathbf{E}}\right|  _{\max}^{2}=2749$ }
\end{figure}

\begin{figure}
\epsfig{file = 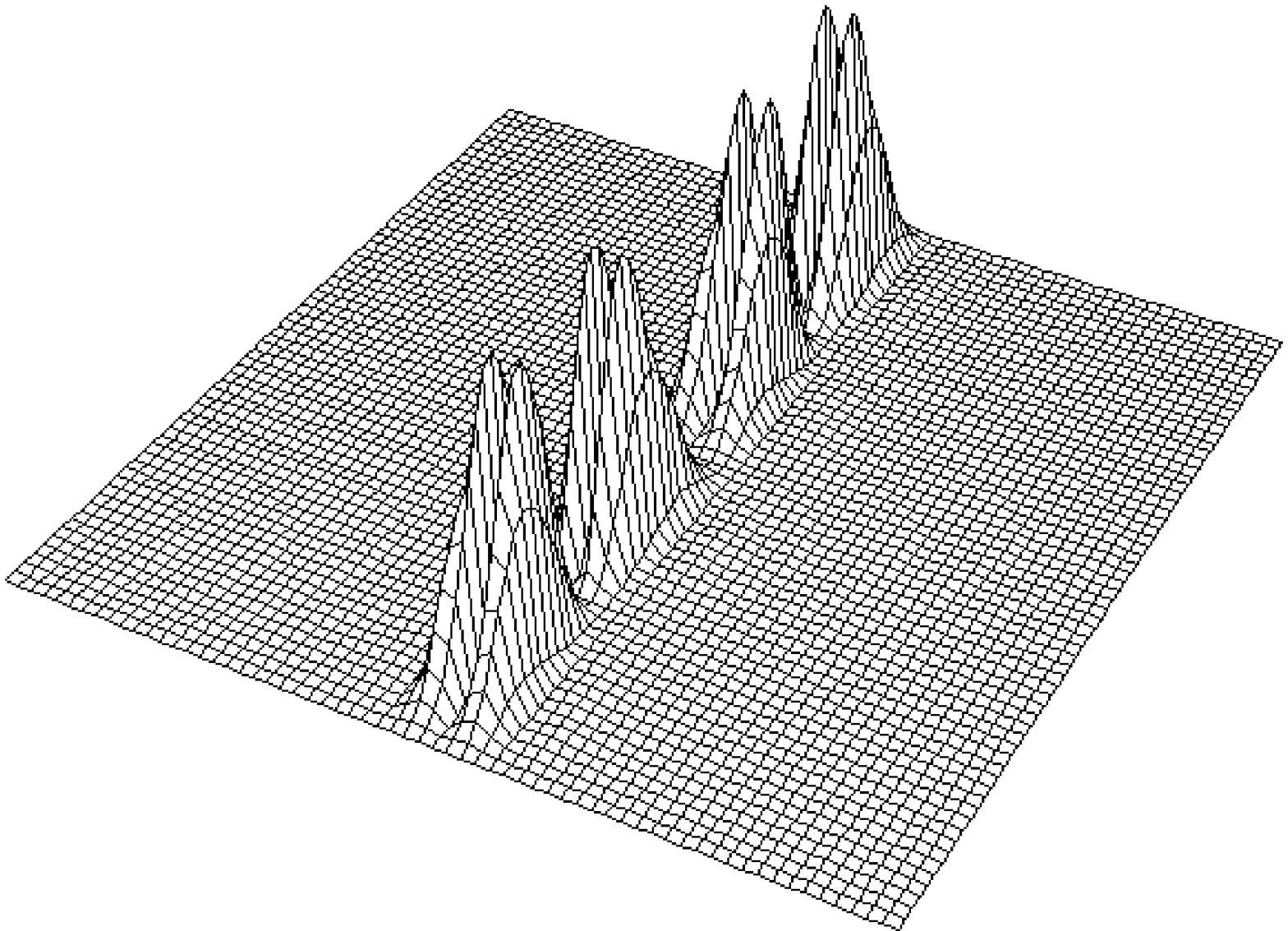, width = 11cm} \caption{$(a)\;\tau= 0.05,\left|
\mathrm{\mathbf{B}} \right|  _{\max}^{2} = 1.171\times10^{ - 22} $ }
\end{figure}

\begin{figure}
\epsfig{file = 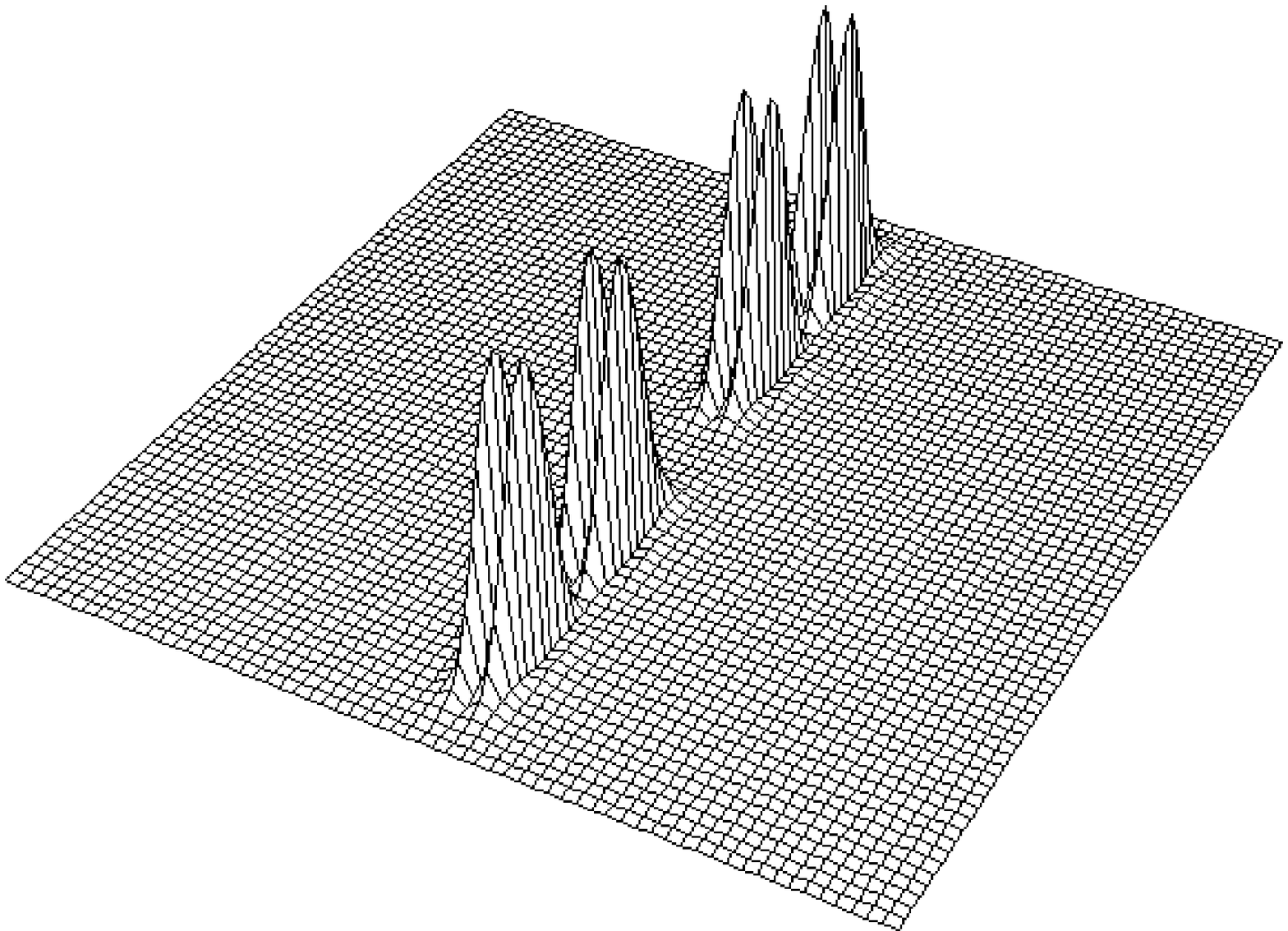, width = 11cm} \caption{$ (b)\;\tau= 0.5,\left|
\mathrm{\mathbf{B}} \right|  _{\max}^{2} = 6.938\times10^{ - 18}$ }
\end{figure}

\begin{figure}
\epsfig{file = 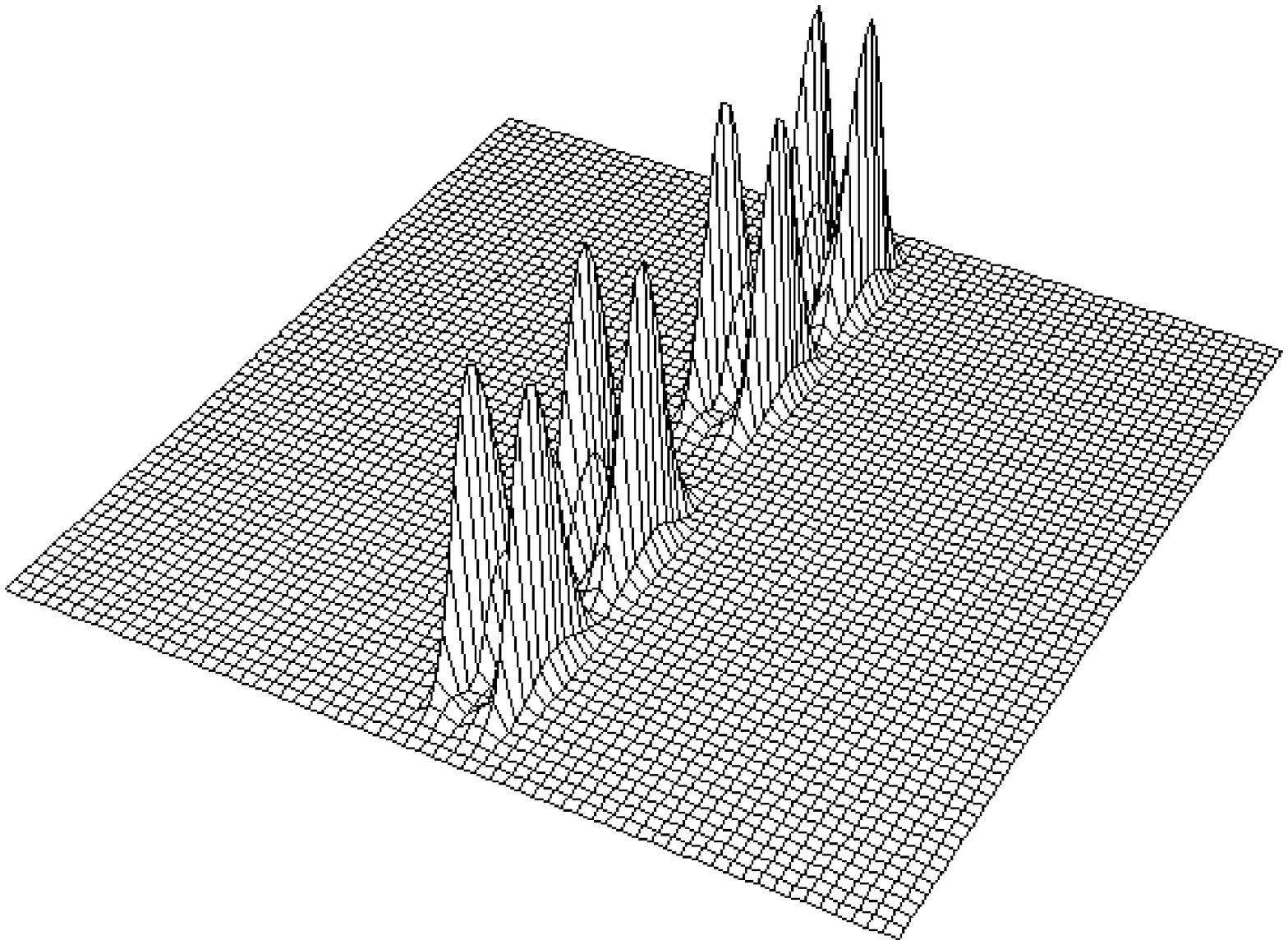, width = 11cm} \caption{$ (c)\;\tau= 1.4,\left|
\mathrm{\mathbf{B}} \right|  _{\max}^{2} = 5.705\times10^{ - 16}$ }
\end{figure}

\begin{figure}
\epsfig{file = 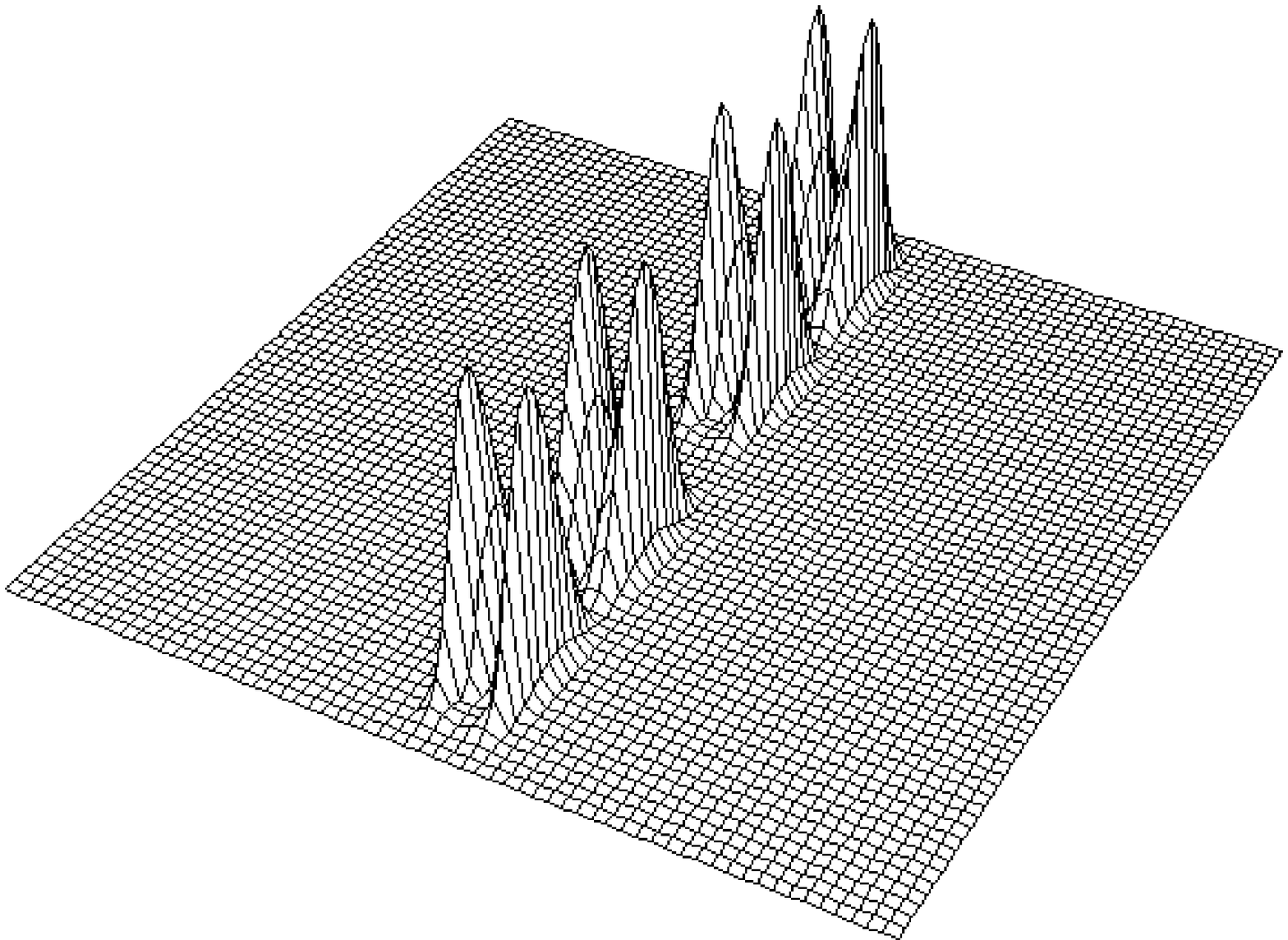, width = 11cm} \caption{$(d)\;\tau= 1.55,\left|
\mathrm{\mathbf{B}} \right|  _{\max}^{2} = 1.154\times10^{ - 15} $ }
\end{figure}

\begin{figure}
\epsfig{file = 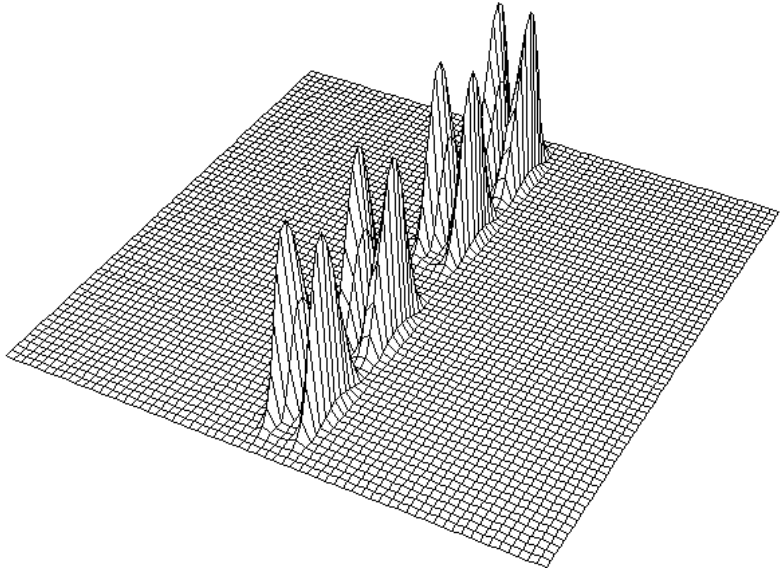, width = 11cm} \caption{$(e)\;\tau=1.6,\left|
\mathrm{\mathbf{B}}\right|  _{\max}^{2}=1.405\times 10^{-15} $ }
\end{figure}

\begin{figure}
\epsfig{file = 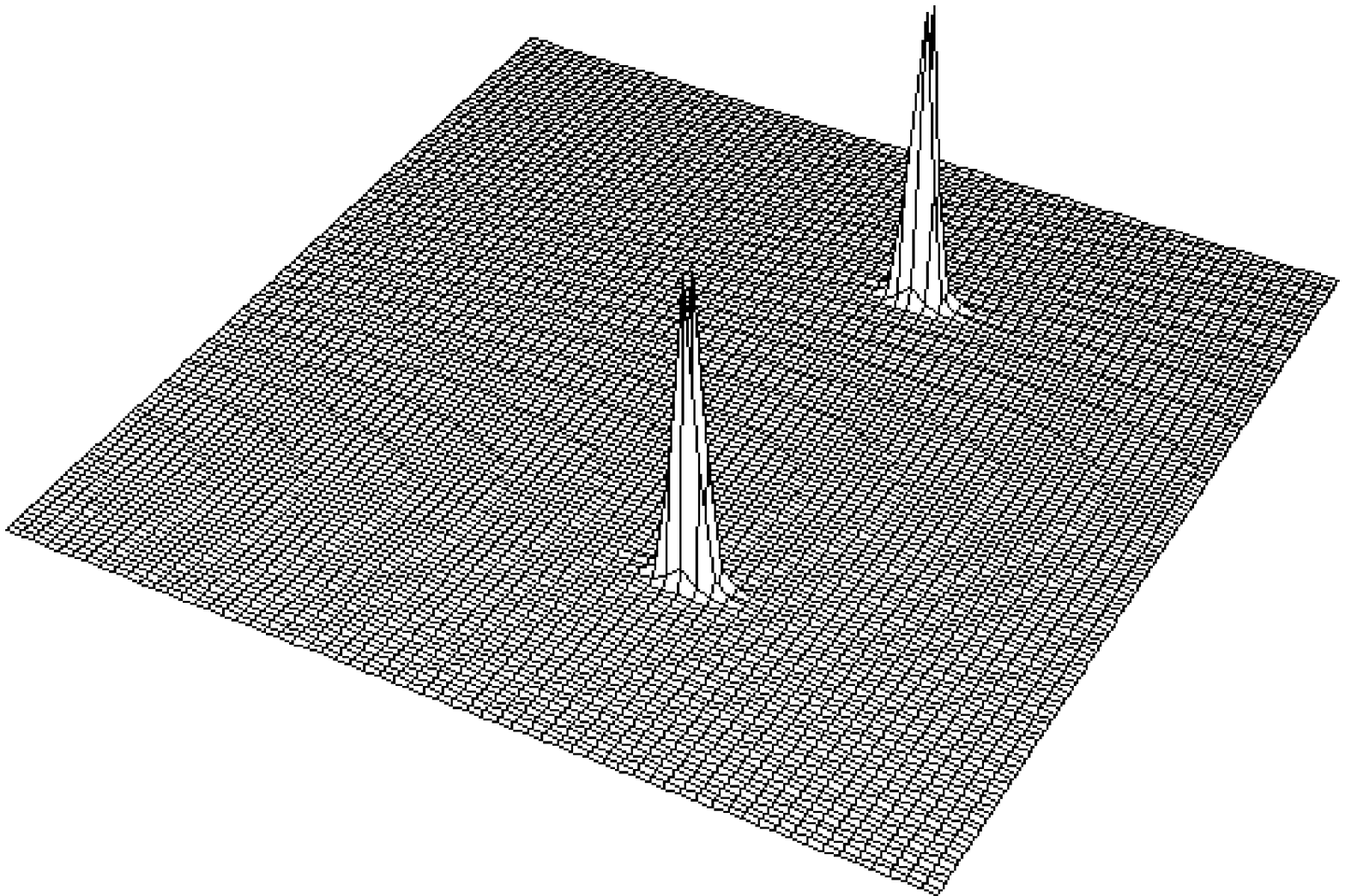, width = 11cm} \caption{$(f)\;\tau=1.65,\left\vert \mathrm{\mathbf{B}}\right\vert _{\max}%
^{2}=8.641\times10^{-9} $ }
\end{figure}

\end{document}